\documentclass[11pt]{article}

\usepackage{algorithm}
\usepackage{algpseudocodex,float}
\usepackage{lipsum}

\usepackage{epsfig}
\usepackage{graphicx}
\usepackage{amsmath}
\usepackage{amsthm}
\usepackage{indentfirst}
\usepackage{cases}
\usepackage{xcolor}
\usepackage{comment}
\usepackage{times}
\usepackage{anysize}
\marginsize{1in}{1in}{1in}{1in}

\usepackage{url}

\usepackage{enumitem}
\setlist[enumerate]{listparindent=\parindent}
\allowdisplaybreaks[4]
\makeatletter

\newtheorem{Proposition}{Proposition}%[section]
\newtheorem{Theorem}{Theorem}%[section]

\begin{document}

\title{On Competitiveness of Dynamic Replication for Distributed Data Access}

\author{
        Tianyu Zuo, Xueyan Tang, Bu Sung Lee, and Jianfei Cai\\
        Nanyang Technological University\\
        Singapore\\
        zuot0001@e.ntu.edu.sg, \{asxytang,~ebslee\}@ntu.edu.sg, jianfei001@e.ntu.edu.sg
}
\date{}
\maketitle

\begin{abstract}
This paper studies an online cost optimization problem for distributed storage and access. The goal is to dynamically create and delete copies of data objects over time at 
geo-distributed servers to serve access requests and minimize the total storage and network cost.
We revisit a recent algorithm in the literature and show that it does \textit{not} have a competitive ratio of $2$ as claimed 
by constructing a counterexample. We further prove that no deterministic online algorithm can achieve a competitive ratio bounded by $2$ for the general cost optimization problem. We develop an online algorithm and prove that it achieves a competitive ratio of $\max\{2, \min\{\gamma, 3\}\}$, where $\gamma$ is the max/min storage cost ratio among all servers. 
Examples are given to confirm the tightness of competitive analysis. We also empirically evaluate algorithms using real object access traces.
\end{abstract}

\section{Introduction}

Replication is an important technique to facilitate data access in distributed systems. The design of replication strategies generally involves trade-offs between the storage and network costs. Creating more copies of a data object (such as 
datasets and videos) can reduce the network cost among various sites to access the object, but it increases the storage cost. Vice versa, maintaining fewer copies of an object can save the storage cost, but it increases the network cost when the object needs to be accessed at sites with no copies. A good replication strategy should strike a balance between the storage and network costs according to distributed access patterns. In addition, the placement of object copies can be dynamically adjusted at run-time following the changes in access patterns. 

In this paper, we study dynamic replication in a distributed system to minimize the total storage and network cost for serving a sequence of data access requests, where both storage and transfer costs are incurred on a pay-as-you-go basis. We focus on the algorithmic challenge of deciding which
data copies to create and hold over time in an online setting where the requests to arise in the future are not known. Our model has useful applications. For example, cloud storage services usually offer multiple sites (datacenters) dispersed at different geographical locations \cite{2017Data}. Storing data at these sites and transferring data among the sites consume resources for cloud providers and incur monetary expenses for cloud customers. Thus, it is valuable to minimize the total resource consumption or monetary cost from their perspectives.
In edge computing \cite{2017Edge}, access providers rent out edge resources to application providers to host their services and serve their users. Since edge platforms often provide pay-as-you-go flexibility \cite{2018Fog}, the application provider needs to mediate data placement across edge servers 
to minimize the 
cost of running the service.

\noindent \textbf{Related work.}
Wei {\em et al.} \cite{5600309} and Gill {\em et al.} \cite{gill2016dynamic} studied the number of replicas needed to meet a given availability requirement in cloud datacenters. Mansouri {\em et al.} \cite{mansouri2018cost} proposed data placement algorithms in cloud storage services that offer two storage tiers characterized by differentiated quality of service and different storage and access costs. 
Dong {\em et al.} \cite{yuan2013highly} and Scouarnec {\em et al.} \cite{scouarnec2014} studied policies for caching data generated from computation in cloud-based systems to balance the trade-off between computation and storage. 
None of the above work, however, addresses the trade-off between storage and transfer costs in distributed storage. There has also been much work on replica placement in networks with consideration of storage and traffic costs \cite{kalpakis2001,tpds2005}. Nevertheless, this line of work normally assumes a fixed storage cost for placing a replica at each node and does not consider pay-as-you-go charging.

Veeravalli \cite{veeravalli2003network} presented a general model for migrating and caching shared data in a network of servers.
Assuming the storage costs of all servers are identical, Bar-Yehuda {\em et al.} \cite{bar2012growing} developed an $O(\log \delta)$-competitive online algorithm where $\delta$ is the normalized diameter of the underlying network.  
Wang {\em et al.} 
\cite{wang2018cost} 
proposed an optimal offline solution by dynamic programming and a $3$-competitive online algorithm under uniform storage costs in servers and transfer costs among servers. 
Later, Wang \textit{et al.} \cite{2021Cost, 2023Cost} further considered a more general case 
where the storage costs of different servers 
are distinct. 
They developed an efficient offline solution based on the concept of shortest path and a $2$-competitive online algorithm. 
They also proved that no deterministic online algorithm can achieve a competitive ratio less than $2$.

Our work is inspired by the above studies and makes substantial algorithmic advancements. 
Our contributions 
are summarized as follows.
\begin{enumerate}[topsep=0pt, partopsep=0pt]
    \item We show that the $2$-competitive claim of Wang \textit{et al.}'s online algorithm in \cite{2021Cost, 2023Cost} is \textit{not true} by constructing a counterexample (Section \ref{sec:counterexample}). The example indicates that the competitive ratio of their algorithm is at least $3$.
    \item We further prove that any deterministic online algorithm has a competitive ratio strictly larger than $2$ for the general cost optimization problem (Section \ref{sec:lowerbound}). 
    \item We propose an online algorithm and prove that it has a tight competitive ratio of $\max\{2, \min\{\gamma, 3\}\}$ (Section \ref{sec:firstalg}), where $\gamma$ is the max/min storage cost ratio among all servers. 
    A major challenge to the competitive analysis is that the exact form of an optimal solution is hard to derive. We develop a novel induction method that exploits the characteristics of an optimal solution.
\end{enumerate}

\section{
Problem Definition}
\label{sec:definition}
We consider a 
system consisting of $n$ 
geo-distributed servers $s_{1},s_{2},...,s_{n}$. 
A data object is hosted in the system and copies of this object can be created and stored in any of the servers.\footnote{
We do not consider any capacity limit on creating data copies in servers, since storage is usually of large and sufficient capacity nowadays.
Hence, we focus on the management of one data object, as different objects can be handled separately.} 
If a copy of the object is stored in server $s_{i}$, it incurs a storage cost of $\mu(s_i)$ per time unit, where $\mu(s_i)$ is called the storage cost rate of $s_{i}$. Different servers may have different storage cost rates. 
We assume that the servers are indexed in ascending order of storage cost rate, i.e., $\mu(s_1) \leq \mu(s_2) \leq \cdots \leq \mu(s_n)$. 
The data object is initially stored in a given server
$s_g \in \{s_{1},s_{2},...,s_{n}\}$. The object can be transferred among different servers. 
The transfer cost of the object between any pair of servers is identical, which is denoted by $\lambda$.

The requests to access the data object may arise at different servers as time goes 
(e.g., due to the computational jobs running at the servers or user requests). When a request arises at a server $s_i$, if $s_{i}$ has a data copy stored locally, it serves the request by its local copy. Otherwise, the object has to be transferred from other servers to $s_{i}$ to serve the request. After serving each local request, $s_{i}$ can choose to store the data copy for some time in anticipation of future requests. This yields a trade-off: it incurs the storage cost but can save the transfer cost if the next local request arises soon. 

We denote the request sequence arising at the servers in the order of time as $R=\{r_{1}, r_{2},...,r_{m}\}$, where $m$ is the number of requests. For each request $r_{j}$, we use $t_{j}$ to denote its arising time and $s[r_{j}]$ to denote the server at which it arises. For simplicity, we assume that the requests arise at distinct times, 
i.e., $t_{1}<t_{2}<...<t_{m}$. For notational convenience, we use $\mu[r_{j}]$ to denote the corresponding storage cost rate of server $s[r_{j}]$, i.e., $\mu[r_{j}] = \mu(s[r_{j}])$. 
Table \ref{tab:notation} summarizes the key notations. 
To facilitate algorithm design and analysis, we
add a dummy request $r_0$ arising at server $s_g$ (the server where the initial copy is located) at time $0$. Note that $r_0$ does not incur any additional cost for
serving the request sequence.

\begin{table}[t]
\centering
\caption{Summary of notations}
\begin{tabular}{|c|l|}
\hline \text { Notation } & \qquad\qquad\qquad\quad\text { Definition } \\
\hline $s_{i}$ & \text { a server in the distributed storage system } \\
$\mu(s_i)$ & \text { the storage cost rate of server $s_{i}$} \\
$r_{j}$ & \text { the $j$-th request to retrieve the data object } \\
$t_{j}$ & \text { the time when request $r_{j}$ arises } \\
$s[r_{j}]$ & \text { the server where request $r_{j}$ arises} \\
$\mu[r_{j}]$ & \text { the storage cost rate of the server where $r_{j}$ arises } \\
$\lambda$ & \text { the transfer cost 
between two servers } \\
\hline
\end{tabular}
\label{tab:notation}
\end{table}

We are interested in developing a \textit{replication strategy} that determines the data copies to create and hold over time and the transfers to carry out in an \textit{online} manner to serve all requests. An essential requirement is that there must be at least one data copy at any time in order to preserve the object. Our objective is to minimize the overall cost of serving a request sequence.

A common metric for evaluating the performance of an online algorithm is the \textit{competitive ratio} \cite{borodin1998}, which is defined as the worst-case ratio between the solution constructed by the online algorithm and an optimal offline solution over all instances of the problem.

\section{Revisiting
Wang \textit{et al.}'s 
Algorithm 
}
\label{sec:counterexample}

Recently, Wang \textit{et al.} \cite{2021Cost, 2023Cost} proposed an online algorithm for the cost optimization problem defined in Section \ref{sec:definition}. They claimed that this algorithm has a competitive ratio of $2$. Unfortunately, we find that the claim is invalid. 

The main idea of Wang \textit{et al.}'s  algorithm is as follows. 
\begin{itemize}
\item For each server $s_{i}$, after serving
a local request (either by a transfer or by the local copy), $s_i$
keeps the data copy for $\frac{\lambda}{\mu(s_i)}$ units of time. Note that the cost of storing the copy in $s_i$ over this period matches the cost of transferring the object. 
\item If a new request arises at $s_i$ in this period, the request is served by the local copy and $s_i$ keeps the copy for another $\frac{\lambda}{\mu(s_i)}$ units of time (starting from the new request). 
\item When the data copy in $s_{1}$ expires (i.e., the server with the lowest storage cost rate), 
the algorithm checks whether $s_{1}$ holds the only copy in the system. If so, $s_{1}$ continues to keep the copy for another $\frac{\lambda}{\mu(s_1)}$ units of time. Otherwise, $s_{1}$ drops its copy. 
\item When the data copy in $s_{i}$ $(i \neq 1)$ expires, the algorithm also checks whether $s_{i}$ holds the only copy. If not, $s_{i}$ drops its copy. If $s_i$ holds the only copy, the algorithm further checks whether $s_{i}$ has kept the copy for $\frac{\lambda}{\mu(s_i)}$ units of time since its most recent local request (i.e., there is no request at $s_i$ for $\frac{\lambda}{\mu(s_i)}$ time). If so, $s_i$ continues to keep the copy for another $\frac{\lambda}{\mu(s_i)}$ units of time. Otherwise, it implies that $s_{i}$ has kept the copy for $\frac{2\lambda}{\mu(s_i)}$ units of time without any local request. In that case, $s_{i}$ transfers the object to $s_{1}$ and drops the local copy.
\end{itemize}

\begin{figure}[tbp]
\centering
\includegraphics[width=8cm]{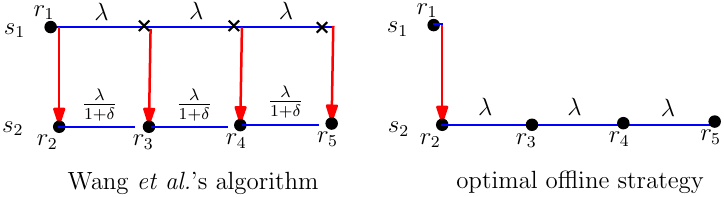}
\caption{\label{j} A counterexample}
\end{figure}

Wang \textit{et al.} \cite{2021Cost, 2023Cost} claimed that this algorithm is $2$-competitive, but we find that it is not true.
We present two counterexamples for the cases of distinct and uniform storage cost rates respectively.
Figure \ref{j} gives a counterexample where two servers $s_{1}$ and $s_{2}$ have storage cost rates $1$ and $1+\delta$ respectively ($\delta > 0$ is a small value).
Each horizontal edge represents a data copy stored in a server from its creation to deletion. Each vertical edge represents a transfer of the object between two servers. 
There are $m \geq 3$ requests in total. Request $r_1$ arises at $s_{1}$ and all subsequent requests arise at $s_{2}$.
The arising times of the requests are $t_{1}=0$, $t_{2}=\epsilon$, $t_{3}=\lambda+\epsilon$, $t_{4}=2\lambda+\epsilon$  
and so on,\footnote{To simplify boundary conditions, Wang \textit{et al.} \cite{2021Cost, 2023Cost} assumed that (1) the object is initially stored in $s_1$ (the server with the lowest storage cost rate)
and (2) the first request arises at $s_1$ at time $0$. Our examples follow these assumptions.} where $\epsilon > 0$ is a small value less than $\lambda-\frac{\lambda}{1+\delta}$ and every two consecutive requests at $s_2$ are $\lambda$ apart.

By Wang \textit{et al.}'s algorithm, after request $r_{1}$, the copy in $s_{1}$ would expire at time $t_1 + \lambda = \lambda$, and after request $r_2$, the copy in $s_2$ would expire at time $t_2 + \frac{\lambda}{1+\delta} = \epsilon + \frac{\lambda}{1+\delta} < \lambda$ which is before $s_1$'s copy expires. Hence, $s_2$ drops its copy when it expires. When $s_1$'s copy expires, since it is the only copy in the system, it is renewed for another $\lambda$ units of time. 
After that, request $r_{3}$ arises at $s_2$. $s_1$ transfers the object to $s_2$ to serve $r_3$, after which the copy in $s_2$ would expire at time $t_3 + \frac{\lambda}{1+\delta} = \lambda + \epsilon + \frac{\lambda}{1+\delta} < 2\lambda$ which is again before $s_1$'s copy expires. Hence, $s_2$ drops its copy when it expires. When $s_1$'s copy expires, it is renewed for another $\lambda$ units of time. This pattern is repeated continuously. As a result, the total cost of serving all requests is at least $(m-2)\cdot3\lambda + \lambda + \epsilon$ (where we include the cost incurred up to the final request $r_m$ only for fair comparison with the optimal offline solution).

In the optimal offline solution, server $s_2$ should always keep a data copy after request $r_2$ is served by a transfer. So, the total cost is $(m-2)\cdot \lambda\cdot (1+\delta) + \lambda + \epsilon$.

As $\delta$ approaches $0$ and $m$ approaches infinity, the online-to-optimal cost ratio is given by
\begin{equation*}
\lim_{m\rightarrow\infty} \lim_{\delta\rightarrow 0} \frac{(m-2)\cdot3\lambda + \lambda + \epsilon}{(m-2)\cdot \lambda\cdot (1+\delta) + \lambda + \epsilon}
= \lim_{m\rightarrow\infty} \frac{(3m-5)\cdot\lambda + \epsilon}{(m-1)\cdot \lambda + \epsilon} 
= 3.
\end{equation*}
This example shows that the competitive ratio of Wang\textit{ et al.}'s algorithm is at least $3$. In fact, given any max/min storage cost ratio greater than $1$ among all servers, an example including two servers with storage cost rates $1$ and $1+\delta$ as above (where $\delta$ is an infinitesimal) can always be constructed. Thus, it is not wise to renew the copy in $s_{1}$ for a fixed time period when it is the only copy.

\begin{figure}[htbp]
\centering
\includegraphics[width=6cm]{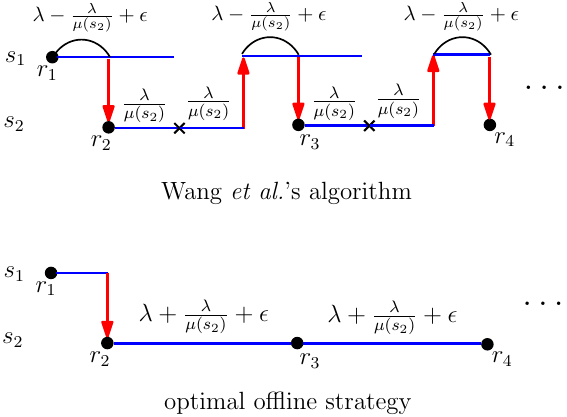}
\caption{\label{wang_new} Another counterexample}
\end{figure}

Figure \ref{wang_new} gives a counterexample where two servers $s_{1}$ and $s_{2}$ have storage cost rates $1$ and $\mu(s_2)$ respectively.
There are $m$ requests. Request $r_1$ arises at $s_{1}$ and all subsequent requests arise at $s_{2}$.
The arising times of the requests are $t_{1}=0$, $t_{2}=\lambda-\frac{\lambda}{\mu(s_2)}+\epsilon$, $t_{3}=2\lambda+2\epsilon$, $t_{4}=3\lambda+\frac{\lambda}{\mu(s_2)}+3\epsilon$
and so on,
where $\epsilon > 0$ is a small value and every two consecutive requests at $s_2$ are $\lambda+\frac{\lambda}{\mu(s_2)}+\epsilon$ apart.

By Wang \textit{et al.}'s algorithm, after request $r_{1}$, the copy in $s_{1}$ would expire at time $t_1 + \lambda = \lambda$, and after request $r_2$, the copy in $s_2$ would expire at time $t_2 + \frac{\lambda}{\mu(s_2)} = \lambda + \epsilon > \lambda$ which is after $s_1$'s copy expires. Hence, $s_1$ drops its copy when it expires. When $s_2$'s copy expires, since it is the only copy in the system, it is renewed for another $\frac{\lambda}{\mu(s_2)}$ time units.
When the renewal expires, $s_2$ transfers the object to $s_1$ and drops the local copy. Then, the copy in $s_1$ would expire at time $2\lambda+\frac{\lambda}{\mu(s_2)}+\epsilon$.
When request $r_{3}$ arises at $s_2$, $s_1$ transfers the object to $s_2$ to serve $r_3$, after which the copy in $s_2$ would expire at time $t_3 + \frac{\lambda}{\mu(s_2)} = 2\lambda + \frac{\lambda}{\mu(s_2)} + 2\epsilon > 2\lambda+\frac{\lambda}{\mu(s_2)} + \epsilon$ which is again after $s_1$'s copy expires. Hence, $s_1$ drops its copy when it expires. When $s_2$'s copy expires, it is renewed for another $\frac{\lambda}{\mu(s_2)}$ time units. When the renewal expires, $s_2$ transfers the object to $s_1$ and drops the local copy. Then, the copy in $s_1$ would expire at time $3\lambda+\frac{2\lambda}{\mu(s_2)}+2\epsilon$.
When request $r_{4}$ arises at $s_2$, $s_1$ transfers the object to $s_2$ to serve $r_4$. 
This pattern is repeated continuously. As a result, the total cost of serving all requests is at least $(m-2)\cdot5\lambda+2\lambda-\frac{\lambda}{\mu(s_2)}+\epsilon$ (where we include the cost incurred up to the final request $r_m$ only for fair comparison with the optimal offline solution).

In the optimal offline solution, server $s_2$ should always keep a data copy after request $r_2$ is served by a transfer. So, the total cost is $(m-2)\cdot \mu(s_2)\cdot(\lambda +\frac{\lambda}{\mu(s_2)}+ \epsilon) + 2\lambda -\frac{\lambda}{\mu(s_2)}+ \epsilon$.

As $\epsilon$ approaches $0$ and $m$ approaches infinity, the online-to-optimal cost ratio is given by
\begin{equation*}
\begin{aligned}
&\lim_{m\rightarrow\infty} \lim_{\epsilon\rightarrow 0} \frac{(m-2)\cdot5\lambda+2\lambda-\frac{\lambda}{\mu(s_2)}+\epsilon}{(m-2)\cdot \mu(s_2)\cdot(\lambda +\frac{\lambda}{\mu(s_2)}+ \epsilon) + 2\lambda -\frac{\lambda}{\mu(s_2)}+ \epsilon} \\ 
  &=\lim_{m\rightarrow\infty} \frac{(5m-8)\cdot\lambda-\frac{\lambda}{\mu(s_2)}}{(m-2)\cdot\left(\mu(s_2)\cdot\lambda+\lambda\right) +2\lambda-\frac{\lambda}{\mu(s_2)}} 
\\&= \frac{5}{\mu(s_2)+1}> 2, \ \text{for all $1\leq\mu(s_2)< \frac{3}{2}$}.
\end{aligned}
\end{equation*}

Setting $\mu(s_2)=1$, this example shows that the competitive ratio of Wang \textit{et al.}'s algorithm is at least $\frac{5}{2}$, even if all servers have the same storage cost rates. Hence, it is not clever to always transfer the object from another server to $s_1$ when the former holds the only copy.

In summary, the examples of Figure \ref{j} and Figure \ref{wang_new} show that Wang \textit{et al.}'s algorithm \cite{2021Cost, 2023Cost} is not $2$-competitive as claimed.

\section{A Lower Bound on Competitiveness}
\label{sec:lowerbound}

In this section, we show that any deterministic online algorithm has a competitive ratio strictly larger than $2$ for the general cost optimization problem defined in Section \ref{sec:definition}.

\begin{Theorem}
\label{the:lowerbound}
For each deterministic online algorithm, there exists an instance of the cost optimization problem such that the cost of the online solution is more than $2$ times the cost of an optimal offline solution.
\end{Theorem}

\begin{proof}
Consider two servers $s_{1}$ and $s_{2}$ with storage cost rates 1 and $\mu > 4$ respectively. Initially (at time $0$), the data object is stored in $s_2$. We define an adversary as follows.

In the online algorithm, if $s_2$ continuously holds the data copy till time $\frac{\lambda}{\mu} + \frac{4\lambda}{\mu^2}$, the adversary makes a request at $s_1$ at time $\frac{\lambda}{\mu} + \frac{4\lambda}{\mu^2}$. Then, the online cost is at least $(\frac{\lambda}{\mu} + \frac{4\lambda}{\mu^2}) \cdot \mu + \lambda$ since a transfer from $s_2$ to $s_1$ is needed to serve the request. 
The optimal offline solution is to transfer the object to $s_1$ at time $0$ and delete the copy in $s_2$, where the cost is $(\frac{\lambda}{\mu} + \frac{4\lambda}{\mu^2}) \cdot 1 + \lambda$. The online-to-optimal cost ratio is given by
\begin{equation*}
\frac{(\frac{\lambda}{\mu} + \frac{4\lambda}{\mu^2}) \cdot \mu + \lambda}{(\frac{\lambda}{\mu} + \frac{4\lambda}{\mu^2}) \cdot 1 + \lambda} = \frac{2\lambda\mu^2 + 4\lambda\mu}{\lambda\mu^2 + \lambda\mu + 4\lambda} > \frac{2\lambda\mu^2 + 4\lambda\mu}{\lambda\mu^2 + 2\lambda\mu} = 2. 
\end{equation*}

In the online algorithm, if $s_2$ stops holding the data copy at some time instant $t < \frac{\lambda}{\mu} + \frac{4\lambda}{\mu^2}$, it must transfer the object to $s_1$ before or at $t$ and $s_1$ will hold the copy after $t$. In this case, the adversary makes a request at $s_2$ at time $t+\epsilon$, where $\epsilon > 0$ is a small value. Then, the object must be transferred from $s_1$ to $s_2$ to serve the request, so the online cost is at least $t \cdot \mu + \epsilon \cdot 1 + 2 \lambda$. The optimal offline solution is to keep the copy at $s_2$ till time $t+\epsilon$ to serve the request, where the cost is $(t+\epsilon) \cdot \mu$. As $\epsilon$ approaches $0$, the online-to-optimal cost ratio is given by
\begin{equation*}
\lim_{\epsilon \rightarrow 0} \frac{t \cdot \mu + \epsilon \cdot 1 + 2 \lambda}{(t+\epsilon) \cdot \mu} 
= \frac{t  \mu + 2 \lambda}{t \mu} 
> 1 + \frac{2\lambda}{\lambda + \frac{4\lambda}{\mu}} > 2. 
\end{equation*}
\end{proof}

\section{
Our Online Algorithm and Analysis}
\label{sec:firstalg}

We would like to design an online algorithm with decent competitiveness. Let us  examine the structure of the problem. At first glance, for every single server, the trade-off between the costs of local storage and inward transfer appears to resemble that between rent and buy in the classical ski-rental problem \cite{1988Com}. If no data copy is stored in the server, we have to pay for the transfer cost at each local request. If a data copy is stored in the server, we have to pay for the storage cost (which however is not a fixed cost for permanent storage and is proportional to the duration of storage). An intuitive idea to achieve $2$-competitiveness is to store a data copy in a server $s_i$ for $\frac{\lambda}{\mu(s_i)}$ units of time after serving each local request. If the next request arises at $s_i$ before this period ends, the request will be served by the local copy and we pay for the storage cost which is optimal. Then, $s_i$ will renew the copy and hold it for another period of $\frac{\lambda}{\mu(s_i)}$. If no request arises at $s_i$ during the period of $\frac{\lambda}{\mu(s_i)}$, $s_i$ will delete the copy and an inward transfer will then be required to serve the next request at $s_i$. In this case, the optimal cost for serving this next request is at least a transfer cost $\lambda$ (not holding a copy in $s_i$) and the cost we pay is at most twice the optimal cost because the storage cost over the $\frac{\lambda}{\mu(s_i)}$ period equals the transfer cost $\lambda$. Nevertheless, if we simply apply this solution to every server, it would not meet the requirement of maintaining at least one data copy at any time, because all copies will be deleted after a sufficiently long silent period without any request. The at-least-one-copy requirement is a key challenge of the problem. 

To meet the at-least-one-copy requirement, we can let the data copy ``follow'' where 
requests arise. If there is only one copy held in a server with a relatively low storage cost rate, we may leave the copy there for any length of time. This can reduce transfers if subsequent requests also arise at this server. On the other hand, if there is only one copy held in a server with a relatively high storage cost rate, we should not keep the copy in this server for too long. Otherwise, the competitive ratio would approach the ratio between the storage cost rates of this server and $s_1$ if there is a long silent period. To address this issue, we can proactively transfer the object to the server $s_1$ with the lowest storage cost. 
Suppose a server $s_i$ keeps a data copy for $\frac{\lambda}{\mu(s_i)}$ units of time after serving a local request and then transfers the object to $s_1$. An intuitive adversarial case is that a new request arises at $s_i$ immediately after the transfer, so the object has to be transferred from $s_1$ back to $s_i$ to serve the request. In this case, we pay a total cost of $3 \lambda$ (the storage cost at $s_i$ is $\lambda$ and each transfer has a cost $\lambda$). An optimal solution is to keep the copy at $s_i$ until the new request arises, which has a cost of $\lambda$. As a result, the online-to-optimal cost ratio is $3$. Thus, we set a threshold of $3 \cdot \mu(s_1)$ to decide whether a server has a high or low storage cost rate. Setting a threshold above $3 \cdot \mu(s_1)$ will make the online-to-optimal cost ratio exceed $3$ when a server with a storage cost rate above $3 \cdot \mu(s_1)$ holds the only copy for a long silent period. 
\subsection{Proposed Online Algorithm} 
Algorithm \ref{alg3} shows the details of our proposed algorithm. 
We use $E_i$ to denote the expiration time of the data copy in server $s_i$, 
use $R_i$ to denote the time of the most recent request arising at $s_i$, 
and use $c$ to denote the number of servers holding data copies. Initially, the data object is stored in a given server $s_{g}$, so $c=1$ (line 2).

\begin{algorithm}[t]
  \caption{Online Algorithm for Dynamic Replication}
  \label{alg3}
  \begin{algorithmic}[1]
  \State /* the data object is initially stored in server $s_g$ */
  \State $\textbf{Initialize:}$ $c \gets 1$; $E_g \gets \frac{\lambda}{\mu(s_g)}$; $E_i \gets -\infty$ for all $i \neq g$; $R_i \gets -\infty$ for all $i$; \Comment{$s_g$ holds the initial copy}
  \If {(a request $r_j$ arises at server $s_{i}$ at time $t_{j}$)}
    \If {$E_i<t_{j}$} \Comment{there is no data copy in $s_{i}$}   
        \State serve $r_{j}$ by a transfer from any other server holding a copy and create a copy in $s_i$;
        \State $E_i \gets t_{j}+\frac{\lambda}{\mu(s_i)}$;
        \State $c \gets c+1$;
    \Else   
        \State serve $r_{j}$ by the local copy in $s_{i}$;
        \State $E_i \gets t_{j}+\frac{\lambda}{\mu(s_i)}$;
    \EndIf
    \State $R_i$ $\gets t_{j}$;
  \EndIf
  \If {($s_{i}$ performs a transfer to another server $s_{k}$ at time $t$ to serve a request at $s_k$)}
    \If {($t-R_i\geq\frac{\lambda}{\mu(s_i)}$)}    
    \Comment{$s_{i}$ holds a special copy}
        \State drop the copy in $s_{i}$;
        \State $E_i \gets t$;
        \State $c \gets c-1$;
    \EndIf
  \EndIf
  \If {(a copy expires in server $s_{i}$ at time $E_i$)}
    \If {$c=1$} \Comment{$s_{i}$ holds the only copy}
        \If {$\mu(s_i) \leq 3\cdot \mu(s_1)$}
            \State $E_i \gets \infty$;
        \Else
            \State $s_i$ performs a transfer to $s_1$;
            \State create a copy in $s_1$;
            \State $E_1 \gets \infty$;
            \State drop the copy in $s_{i}$;
        \EndIf
    \Else
        \State drop the copy in $s_{i}$;
        \State $c \gets c-1$;
    \EndIf
  \EndIf
  \end{algorithmic}
\end{algorithm}

Due to the trade-off between the storage and transfer costs, we let each server $s_{i}$ keep the data copy for $\frac{\lambda}{\mu(s_i)}$ units of time after serving every local request (lines 3-11).\footnote{With the dummy request $r_0$ arising at server $s_g$ (the server where the initial copy is located) at time $0$, $s_g$ would hold a data copy for $\frac{\lambda}{\mu(s_g)}$ time at the beginning (line 2).}
This is because the storage cost will be lower than the transfer cost if the next local request arises within this period.
When the data copy in a server $s_{i}$ expires (line 17), if $s_{i}$ does not hold the only copy in the system, it drops the copy (lines 26-28). If $s_{i}$ holds the only copy and has a storage cost rate $\mu(s_i) \leq 3\cdot \mu(s_1)$, it continues to keep the copy (lines 19-20). If $s_{i}$ holds the only copy and has a storage cost rate $\mu(s_i) > 3\cdot \mu(s_1)$, it transfers the data object to server $s_1$ and then $s_1$ continues to keep the copy (lines 21-25).  
In the latter two cases, subsequently, 
once $s_i$ or $s_1$ transfers the data object to another server (for serving a request at that server), it drops its local copy right after the transfer (lines 12-16). Then, a copy will be created at the server receiving the transfer (line 5), so the requirement of maintaining at least one data copy at any time is still satisfied. 
Note that since $s_i$ or $s_1$ drops its copy immediately upon an outward transfer, we can simply check whether the duration since the most recent request at $s_i$ or $s_1$ is longer than $\frac{\lambda}{\mu(s_i)}$ or $\frac{\lambda}{\mu(s_1)}$ to identify the above two cases
(line 13). 

\begin{figure}[tbp]
\centering
\includegraphics[width=8.5cm]{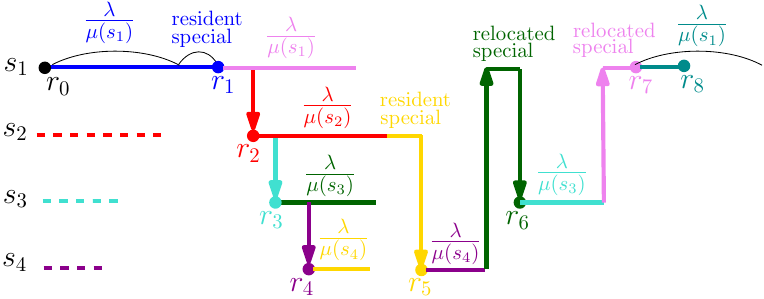}
\caption{An example of our algorithm}
\label{costalloc}
\end{figure}

Figure \ref{costalloc} shows an example of our algorithm, where $\mu(s_2) \leq 3\cdot \mu(s_1)$, $\mu(s_3) > 3\cdot \mu(s_1)$ and $\mu(s_4) > 3\cdot \mu(s_1)$. To facilitate analysis, we categorize data copies. After serving a local request, a server $s_i$ keeps a data copy for up to $\frac{\lambda}{\mu(s_i)}$ time. We refer to the data copy during this $\frac{\lambda}{\mu(s_i)}$ period as the \textit{regular copy}. Upon the expiration of this time period, $s_i$ would continue to keep the data copy if it is the only copy in the system and $\mu(s_i) \leq 3\cdot \mu(s_1)$. In this case, we refer to the data copy beyond the $\frac{\lambda}{\mu(s_i)}$ period as the \textit{resident special copy}. On the other hand, if $s_i$'s expiring copy is the only copy and $\mu(s_i) > 3\cdot \mu(s_1)$, $s_i$ would transfer the data object to $s_1$ which will then keep a data copy. In this case, we refer to the data copy created at $s_1$ as the \textit{relocated special copy}.  By the algorithm definition, it is easy to infer the following property. 

\begin{Proposition}
The storage periods of any two special copies do not overlap. Moreover, the storage period of any special copy does not overlap with that of any regular copy.
\label{pro-2}
\end{Proposition}

Applying our algorithm to the example of Figure \ref{j}, server $s_2$ would always keep a data copy after request $r_3$ is served by a transfer, so the online cost is $(m-3)\cdot \lambda\cdot (1+\delta) + 4\lambda + \epsilon$, where $4\lambda$ is the cost incurred between $r_2$ and $r_3$. Then, the ratio between the online cost and the optimal offline cost is bounded by $2$ for any $m \geq 3$ and it approaches $1$ as $m\rightarrow\infty$. Similarly, in the example of Figure \ref{wang_new},
if $1\leq\mu(s_2)< \frac{3}{2}$, server $s_2$ would always keep a data copy after request $r_2$ is served by a transfer, so the online cost is $(m-2)\cdot \mu(s_2)\cdot(\lambda +\frac{\lambda}{\mu(s_2)}+ \epsilon) + 2\lambda$, 
where the additive term $2\lambda$ refers to the transfer cost to serve $r_2$ and the storage cost of the copy in $s_1$ after $r_1$. Then, the ratio between the online cost and the optimal offline cost is bounded by $2$ for any $m\geq 2$ and it approaches $1$ as $m\rightarrow\infty$.

Next, we will analyze Algorithm \ref{alg3} and show that it is $\max\{2, \min\{\gamma, 3\}\}$-competitive, where $\gamma = \frac{\max_i \mu(s_i)}{\min_i \mu(s_i)}$ is the max/min ratio of the storage cost rates of all servers. 

\subsection{Preliminaries: Allocation of Online Cost} \label{1a}
Our approach to competitive analysis is induction. 
Given a request sequence, we shall prove that the ratio between the online cost and the optimal offline cost to serve any prefix of the request sequence is bounded by $\max\{2, \min\{\gamma, 3\}\}$. To enable cost computation over a request subsequence,
we first present a method to allocate the online cost to individual requests. 

\begin{figure*}[htbp]
\centering
\includegraphics[width=17cm]{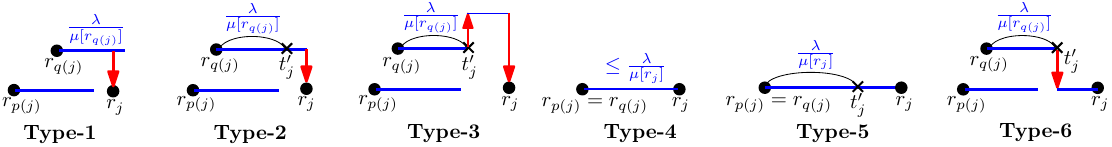}
\caption{Illustration of different request types in our online algorithm}
\label{type1234} 
\end{figure*}

To facilitate presentation, for a request $r_j$, we define $r_{p(j)}$ as the preceding request of $r_{j}$ arising at the same server as $r_j$ does. As illustrated in Figure \ref{type1234}, we categorize all the requests in a sequence into six types based on how they are served by our online algorithm. Recall that a request is served by either a transfer or a local copy.

If a request $r_j$ is served by a transfer from another server $s$, we categorize $r_j$ by the copy type held by $s$ at the time of $r_j$. If the copy held by $s$ is a regular copy, 
$r_j$ is called a \textbf{Type-1} request. If the copy held by $s$ 
is a resident special copy, 
$r_j$ is called a \textbf{Type-2} request. In these two cases, by the algorithm definition, $s$ must have kept the data copy since the most recent request at $s$ before $r_j$ arises. We denote this most recent request by $r_{q(j)}$ and then $s = s[r_{q(j)}]$. For each \textbf{Type-2} request $r_{j}$, we define $t'_{j}$ as the time instant when the data copy in $s[r_{q(j)}]$ switches from a regular copy to a special copy, i.e., $t'_{j} = t_{q(j)} + \frac{\lambda}{\mu[r_{q(j)}]}$.
If the copy held by $s$ is a relocated special copy, $r_j$ is called a \textbf{Type-3} request. In this case, by the algorithm definition, $s = s_1$ and the copy in $s_1$ must be created due to a transfer from another server who keeps a data copy since its most recent request. We also denote this most recent request by $r_{q(j)}$ and its server by $s[r_{q(j)}]$. 
We define $t'_{j}$ as the time instant when $s[r_{q(j)}]$ transfers the data object to $s_1$, i.e., $t'_{j} = t_{q(j)} + \frac{\lambda}{\mu[r_{q(j)}]}$. In this way, the notation $r_{q(j)}$ consistently refers to the request that provides the data copy to serve $r_j$ by our online algorithm, and the notation $t'_{j}$ consistently refers to the time instant when the regular copy following $r_{q(j)}$ switches to a special copy if so.

If a request $r_j$ is served by the local copy in the server $s[r_j]$, we categorize $r_j$ by the copy type held by $s[r_j]$ at the time of $r_j$. If the copy held by $s[r_j]$ is a regular copy, 
$r_j$ is called a \textbf{Type-4} request. If the copy held by $s[r_j]$ is a resident special copy, 
$r_j$ is called a \textbf{Type-5} request. 
In these two cases, by the algorithm definition, $s[r_j]$ must have kept the data copy since the preceding request $r_{p(j)}$ at $s[r_j]$. 
For notational convenience, for each \textbf{Type-4/5} request $r_{j}$, we define $r_{q(j)} = r_{p(j)}$. 
For each \textbf{Type-5} request $r_j$, we also define $t'_{j}$ as the time instant when the data copy in $s[r_{j}]$ switches from a regular copy to a special copy, i.e., $t'_{j} = t_{q(j)} + \frac{\lambda}{\mu[r_{q(j)}]}$. 
If the copy held by $s[r_j]$ is a relocated special copy, $r_j$ is called a \textbf{Type-6} request. In this case, by the algorithm definition, $s[r_j] = s_1$ and the copy in $s_1$ must be created due to a transfer from another server who keeps a data copy since its most recent request. We again denote this most recent request by $r_{q(j)}$ and its server by $s[r_{q(j)}]$. We define $t'_{j}$ as the time instant when $s[r_{q(j)}]$ transfers the data object to $s_1$, i.e., $t'_{j} = t_{q(j)} + \frac{\lambda}{\mu[r_{q(j)}]}$ is the time when the regular copy following $r_{q(j)}$ expires.

Since there are only three types of data copies by running our algorithm, the above request categorization is apparently complete.

We make an important observation on the duration between a request $r_j$ and the preceding request $r_{p(j)}$ at the same server. 
\begin{Proposition}
If $r_j$ is not a \textbf{Type-4} request, then $t_{j}-t_{p(j)} > \frac{\lambda}{\mu[r_{j}]}$. If $r_j$ is a \textbf{Type-4} request, then $t_{j}-t_{p(j)} \leq \frac{\lambda}{\mu[r_{j}]}$.
\label{pro-3}
\end{Proposition}
\begin{proof}
If $r_{j}$ is a \textbf{Type-1/2/3} request, since it is served by a transfer from some other server, the data copy in $s[r_j]$ after $r_{p(j)}$ has been dropped at the time of $r_j$. Thus, we must have $t_{j}-t_{p(j)} > \frac{\lambda}{\mu[r_{j}]}$.
If $r_j$ is a \textbf{Type-4/5/6} request, the duration between $t_{p(j)}$ and $t_j$ is clear by definition.  
\end{proof}

The total online cost of Algorithm \ref{alg3} consists of three parts: (1) the storage cost of regular copies; (2) the storage cost of special copies; and (3) the cost of transfers. We allocate them to individual requests as follows.
\begin{itemize}
\item The storage cost of a regular copy after request $r_{p(j)}$ is allocated to request $r_j$. Note that this storage cost is $\lambda$, unless $r_j$ is a \textbf{Type-4} request.
\item The storage cost of a special copy is allocated to the request that it serves. That is, the storage cost from time $t'_{j}$ to $t_j$ is allocated to $r_j$ in the illustrations of \textbf{Type-2/3/5/6} requests in Figure \ref{type1234}.
\item The cost $\lambda$ of a transfer to serve a request $r_j$ is allocated to $r_j$. This refers to the transfer at time $t_j$ in the illustrations of \textbf{Type-1/2/3} requests in Figure \ref{type1234}. 
\item The cost $\lambda$ of a transfer to create a relocated special copy is allocated to the request that it serves. That is, the transfer cost at time $t'_{j}$ is allocated to $r_j$ in the illustrations of \textbf{Type-3/6} requests in Figure \ref{type1234}. 
\end{itemize}

In the example of Figure \ref{costalloc}, 
each request and its allocated cost are marked in the same color. 
$r_{2}$, $r_{3}$ and $r_{4}$ are \textbf{Type-1} requests, $r_{5}$ is a \textbf{Type-2} request, $r_{6}$ is a \textbf{Type-3} request, $r_{8}$ is a \textbf{Type-4} request, $r_{1}$ is a \textbf{Type-5} request, and $r_{7}$ is a \textbf{Type-6} request.

The following proposition summarizes the cost allocation.
\begin{Proposition}
\label{costsummary}
The online cost allocated to a request $r_j$, based on its type, is given by
\begin{itemize}
\item \textbf{Type-1:} 
$2\lambda$; 
\item \textbf{Type-2:} 
$2\lambda+\mu[r_{q(j)}]\cdot\big(t_{j}-t'_{j}\big) \leq 2\lambda+\min\{\gamma, 3\} \cdot \mu(s_1)\cdot\big(t_{j}-t'_{j}\big)$;
\item \textbf{Type-3:} 
$3\lambda+\mu(s_1)\cdot\big(t_{j}-t'_{j}\big)$;
\item \textbf{Type-4:} 
$\mu[r_{j}]\cdot\big(t_{j}-t_{p(j)}\big)$; 
\item \textbf{Type-5:} 
$\lambda+\mu[r_{q(j)}]\cdot\big(t_{j}-t'_{j}\big) \leq \lambda+\min\{\gamma, 3\} \cdot \mu(s_1)\cdot\big(t_{j}-t'_{j}\big)$. 
\item \textbf{Type-6:} 
$2\lambda+\mu(s_1)\cdot\big(t_{j}-t'_{j}\big)$;
\end{itemize}
Note that the first term in the allocated cost of a \textbf{Type-1/2/3/5/6} request is always bounded by $\max\{2, \min\{\gamma, 3\}\}\cdot\lambda$, since a \textbf{Type-3} request can exist only if the max/min storage cost ratio $\gamma > 3$.
In addition, by the definition of Algorithm \ref{alg3}, resident special copies are stored at servers with storage cost rates at most $\min\{\gamma, 3\} \cdot \mu(s_1)$, which brings the bound on the allocated cost of a \textbf{Type-2/5} request.
\end{Proposition}

There are some special considerations to note in the cost allocation of regular copies. Let $r_m$ denote the final request in the request sequence. 
After serving $r_m$, a regular copy is created in server $s[r_m]$. Among this regular copy and all other regular copies that exist 
after $r_m$, the copy expiring the latest would switch to a special copy and stay infinitely.
In our analysis, we shall not account for the cost of the regular copy created in server $s[r_m]$ after $r_m$ and the special copy which stays infinitely. The rationale is that the storage periods of these two copies do not overlap and they are both entirely beyond 
$r_m$. These two copies are considered to be in existence for 
meeting the at-least-one-copy requirement 
beyond $r_m$. 
In an optimal offline strategy, no copy will need to be stored beyond $r_m$. Thus, to focus on 
a finite time horizon, we do not account for the cost of the aforementioned two copies by the online algorithm.\footnote{We remark that it would not affect the correctness of our competitive analysis even if we insist in considering an infinite time horizon. Please refer to Appendix \ref{infinite} for an elaboration.}

If there are $n$ servers ever holding copies in serving a request sequence,
there will be a storage cost of $(n-1)\lambda$ for the regular copies after the last request at each server (except the server where $r_m$ arises), which has not been allocated in the earlier discussion.
Note that 
serving the first request at each server (except $s_{g}$ with the initial copy) must involve a transfer. 
Thus, such first requests cannot be \textbf{Type-4/5} requests. Since there are $(n-1)$ such first requests in total, we allocate the storage cost of $(n-1)\lambda$ to these $(n-1)$ first requests (a cost of $\lambda$ for each request). In this way, the first request $r_{j}$ of each server (except $s_{g}$) is allocated a total cost of $2\lambda$ if it is served by a transfer from a regular copy (\textbf{Type-1} request), 
a total cost of $2\lambda + \mu[r_{q(j)}]\cdot\big(t_{j}-t'_{j}\big)$ if it is served by a transfer from a resident special copy (\textbf{Type-2} request), a total cost of $3\lambda + \mu(s_1)\cdot\big(t_{j}-t'_{j}\big)$ if it is served by a transfer from a relocated special copy (\textbf{Type-3} request), or a total cost of $2\lambda + \mu(s_1)\cdot\big(t_{j}-t'_{j}\big)$ if it is served by the local copy which is a relocated special copy (\textbf{Type-6} request).  
This is then consistent with the cost allocations of other \textbf{Type-1/2/3/6} requests as given in Proposition \ref{costsummary}.

In the example of Figure \ref{costalloc}, $r_{2}$, $r_{3}$ and $r_{4}$ are the first requests arising at servers $s_{2}$, $s_{3}$ and $s_{4}$ respectively, and they are all \textbf{Type-1} requests. On the other hand, we do not account for the regular copy after the final request $r_8$ which arises at $s_1$. There are three regular copies (each of storage cost $\lambda$) after the last requests at the other servers. This storage cost of $3\lambda$ is allocated to $r_{2}$, $r_{3}$ and $r_{4}$, with $\lambda$ for each (dashed lines). As a result, $r_{2}$, $r_{3}$ and $r_{4}$ are each allocated a total cost of $2\lambda$.

It is easy to verify that 
the sum of the costs allocated to all requests is equal to the total online cost and the dummy request $r_0$ is not allocated any cost.

\subsection{Competitive Analysis by Induction} \label{induction}

Given a request sequence $R=\left\langle r_{1},r_{2},...,r_{m} \right\rangle$, we use $\textbf{Online$(i,j)$}$ to denote the total online cost allocated to the requests from $r_{i}$ to $r_{j}$ ($1\leq i\leq j \leq m$). Since each request is allocated a separate partition of the online cost, it apparently holds that $\textbf{Online$(i,j)$} = \textbf{Online$(1,j)$} - \textbf{Online$(1,i)$}$. Note that all $\textbf{Online$(i,j)$}$'s refer to portions of the cost produced by running Algorithm \ref{alg3} on the request sequence $R$. 

In addition, we use $\textbf{OPT$(1,i)$}$ to denote the optimal offline cost of serving the request subsequence $\left\langle r_{1},r_{2},...,r_{i} \right\rangle$, and define $\textbf{OPT$(i,j)$} = \textbf{OPT$(1,j)$} - \textbf{OPT$(1,i)$}$ for any $1\leq i\leq j \leq m$. Note that each $\textbf{OPT$(i,j)$}$ refers to the cost of the optimal strategy for serving a distinct subsequence of $R$. Hence, different $\textbf{OPT$(i,j)$}$'s refer to different strategies.

\subsubsection{Base Case} 

In the base case, we consider only the first request $r_1$ and prove that $\frac{\textbf{Online$(1,1)$}}{\textbf{OPT$(1,1)$}}\leq \max\{2, \min\{\gamma, 3\}\}$.
Remember that the initial copy is located at server $s_g$. 

If $r_1$ is a \textbf{Type-1} request, it is served by a transfer from a regular copy at $s_g$, so the online cost is $2\lambda$. The optimal offline strategy also requires a transfer to serve $r_1$ and hence has a cost at least $\lambda$. Thus, the online-to-optimal cost ratio is bounded by $2$. 

If $r_1$ is a \textbf{Type-2} request, it is served by a transfer from a resident special copy at $s_g$, so we must have $\mu(s_g) \leq 3\cdot \mu(s_1)$ and $t_1 > \frac{\lambda}{\mu(s_g)}$ by the algorithm definition. The online cost is $\mu(s_g)\cdot t_1 + \lambda$.
The optimal offline strategy is either the same, or it transfers the object at time $0$ to server $s[r_1]$ and keeps the copy at $s[r_1]$ till $r_1$ (cost is $\mu[r_1]\cdot t_1 + \lambda$), or it transfers the object at time $0$ to the server $s_1$ with the lowest storage cost rate and transfers it to $s[r_1]$ at $r_1$ (cost is $\mu(s_1)\cdot t_1 + 2\lambda$). In the second case, the online-to-optimal cost ratio is $\frac{\mu(s_g)\cdot t_1 + \lambda}{\mu[r_1]\cdot t_1 + \lambda} < \gamma$,
where $\gamma$ is the max/min storage cost ratio of all servers. We also have
\begin{equation*}
\frac{\mu(s_g)\cdot t_1 + \lambda}{\mu[r_1]\cdot t_1 + \lambda} \leq \frac{3\cdot \mu(s_1)\cdot t_1 + \lambda}{\mu[r_1]\cdot t_1 + \lambda} < 3,
\end{equation*}
because server $s_1$ has the lowest storage cost rate. Thus, the online-to-optimal cost ratio is bounded by $\min\{\gamma, 3\}$. In the third case, the online-to-optimal cost ratio is $\frac{\mu(s_g)\cdot t_1 + \lambda}{\mu(s_1)\cdot t_1 + 2\lambda} < \min\{\gamma, 3\}$.

If $r_1$ is a \textbf{Type-3} request, it is served by a transfer from a relocated special copy at $s_1$, so we must have $\mu(s_g) > 3\cdot \mu(s_1)$ and $t_1 > \frac{\lambda}{\mu(s_g)}$ by the algorithm definition. The online cost is $\mu(s_1)\cdot (t_1 - \frac{\lambda}{\mu(s_g)}) + 3\lambda$. The optimal offline strategy 
either keeps the copy at $s_g$ till $r_1$ (cost is at least $\mu(s_g)\cdot t_1$), or it transfers the object at time $0$ to server $s[r_1]$ and keeps the copy at $s[r_1]$ till $r_1$ (cost is $\mu[r_1]\cdot t_1 + \lambda$), or transfers the object at time $0$ to server $s_1$ and transfers it back to $s_g$ at $r_1$ (cost is $\mu(s_1)\cdot t_1 + 2 \lambda$). 
In the first case, the online-to-optimal cost ratio is
\begin{equation*}
\begin{aligned}
\frac{\mu(s_1)\cdot (t_1 - \frac{\lambda}{\mu(s_g)}) + 3\lambda}{\mu(s_g)\cdot t_1} & < \frac{\mu(s_1)}{\mu(s_g)} + \frac{3\lambda - \lambda\cdot\frac{\mu(s_1)}{\mu(s_g)}}{\lambda} 
\\
& = 3 = \min\{\gamma, 3\}.
\end{aligned}
\end{equation*}
In the second case, the online-to-optimal cost ratio is 
\begin{equation*}
\frac{\mu(s_1)\cdot (t_1 - \frac{\lambda}{\mu(s_g)}) + 3\lambda}{\mu[r_1]\cdot t_1 + \lambda} < 3 = \min\{\gamma, 3\}.
\end{equation*}
In the third case, the online-to-optimal cost ratio is 
\begin{equation*}
\frac{\mu(s_1)\cdot (t_1 - \frac{\lambda}{\mu(s_g)}) + 3\lambda}{\mu(s_1)\cdot t_1 + 2 \lambda} < \frac{\mu(s_1)\cdot t_1 + 3\lambda}{\mu(s_1)\cdot t_1 + 2 \lambda} < \frac{3}{2}.
\end{equation*}

If $r_1$ is a \textbf{Type-4} request, it is served locally by a regular copy, so $r_1$ is at $s_g$ and the online cost is $\mu(s_g)\cdot t_1 \leq \lambda$. In this case, the optimal offline strategy must be the same. 

If $r_1$ is a \textbf{Type-5} request, it is served locally by a residential special copy, so $r_1$ is at $s_g$ and we have $\mu(s_g) \leq 3\cdot \mu(s_1)$ by the algorithm definition. The online cost is $\mu(s_g)\cdot t_1$. The optimal offline strategy is either the same or it transfers the object at time $0$ to server $s_1$ and transfers it back to $s_g$ at $r_1$ (cost is $\mu(s_1)\cdot t_1 + 2 \lambda$). In the latter case, the online-to-optimal cost ratio is 
\begin{equation*}
\frac{\mu(s_g)\cdot t_1}{\mu(s_1)\cdot t_1 + 2 \lambda} < \frac{\mu(s_g)}{\mu(s_1)} \leq \min\{\gamma, 3\}.
\end{equation*}

If $r_1$ is a \textbf{Type-6} request, it is served locally by a relocated special copy, so $r_1$ is at $s_1$ and we have $\mu(s_g) > 3\cdot \mu(s_1)$ and $t_1 > \frac{\lambda}{\mu(s_g)}$ by the algorithm definition. The online cost is $\mu(s_1)\cdot(t_1 - \frac{\lambda}{\mu(s_g)})+2\lambda$. The optimal offline strategy is to transfer the object to $s_1$ at time $0$ (cost is $\mu(s_1)\cdot t_1+\lambda$). Thus, the online-to-optimal cost ratio is 
\begin{equation*}
\frac{\mu(s_1)\cdot(t_1 - \frac{\lambda}{\mu(s_g)})+2\lambda}{\mu(s_1)\cdot t_1+\lambda} < \frac{\mu(s_1)\cdot t_1+2\lambda}{\mu(s_1)\cdot t_1+\lambda} < 2.
\end{equation*}

In summary, the ratio between the online cost and the optimal offline cost of $r_1$ is bounded by $\max\{2, \min\{\gamma, 3\}\}$.

\subsubsection{Induction Step} 

Suppose that for any $j\in\left\{1,2,...,i-1\right\}$, the ratio between the online cost and the optimal offline cost for the first $j$ requests is bounded by $\max\{2, \min\{\gamma, 3\}\}$, i.e., $\frac{\textbf{Online$(1,j)$}}{\textbf{OPT$(1,j)$}}\leq \max\{2, \min\{\gamma, 3\}\}$. 
In the induction step, we show that the ratio is also bounded by $\max\{2, \min\{\gamma, 3\}\}$ for the first $i$ requests, i.e., $\frac{\textbf{Online$(1,i)$}}{\textbf{OPT$(1,i)$}}\leq \max\{2, \min\{\gamma, 3\}\}$. 

A major challenge here is that 
the exact form of an optimal offline strategy 
is not straightforward to derive. We complete the induction step by exploiting
two important characteristics of an optimal offline strategy presented below. 
Recall that $r_{p(i)}$ denotes the preceding request of $r_{i}$ arising at the same server as $r_i$ does.  

First, if two successive requests at the same server are sufficiently close in time, the server should hold a copy between them.

\begin{Proposition} \label{prop5}
There exists an optimal offline 
strategy in which for each request $r_i$, 
if $\mu[r_i] \cdot (t_i - t_{p(i)}) \leq \lambda$, server $s[r_i]$ holds a copy throughout
the period $(t_{p(i)}, t_i)$.
\end{Proposition}
\begin{proof}
If server $s[r_i]$ does not hold a copy all the time from $t_{p(i)}$ to $t_i$,
server $s[r_i]$ must receive a transfer sometime in the period $(t_{p(i)}, t_i)$ in order to serve request $r_i$,
where the transfer cost incurred is $\lambda$. 
Since $\mu[r_i] \cdot (t_i - t_{p(i)}) \leq \lambda$, the transfer can be replaced by storing a copy from $t_{p(i)}$ to $t_i$ in $s[r_i]$ to either save cost (which contradicts the cost optimality) or maintain the same total cost. 
\end{proof}

Second, each transfer is performed at the time when some request arises.

\begin{Proposition}
There exists an optimal offline strategy with
the characteristic in Proposition \ref{prop5} and that 
there is a request at either the source server or the destination server of each transfer.
\label{pro-1}
\end{Proposition}

The main idea to prove Proposition \ref{pro-1} is that if there is no request at the source and destination servers, we can always advance or delay the transfer to save or maintain cost based on the relative storage cost rates of the source and destination servers. The complete proof is provided in 
Appendix \ref{proofreqatsrcdes}.

By Proposition \ref{pro-1}, the possible ways of serving each request in an optimal offline strategy are largely restricted.
We can divide all requests into the following four categories based on how they are served in the optimal offline strategy.

\begin{figure}[tbp]
\centering
\includegraphics[width=8cm]{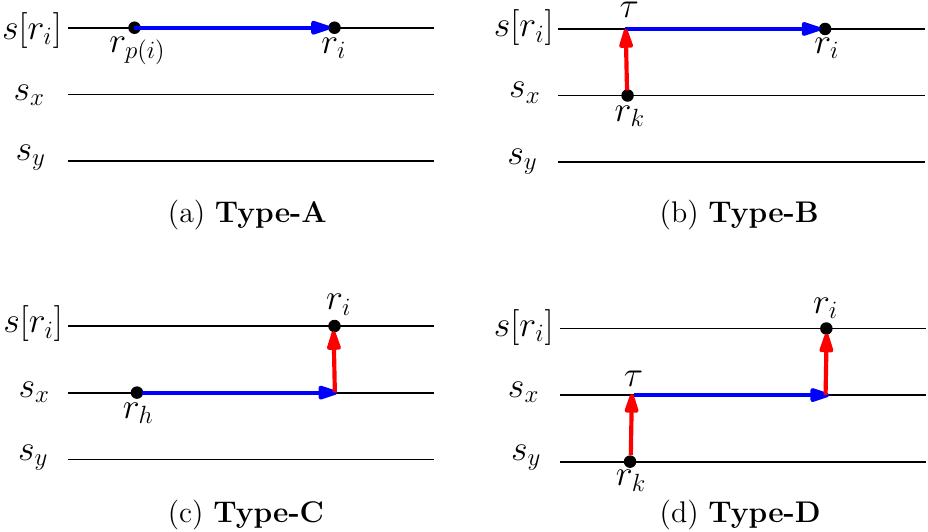}
\caption{\label{12} Four types of requests in an optimal offline strategy}
\end{figure}

\medskip
\noindent {\bf Type-A Requests.} A request $r_i$ is served by the local copy in server $s[r_i]$ and this local copy was created no later than the preceding request $r_{p(i)}$ (see Figure
\ref{12}(a)). 

\medskip
\noindent {\bf Type-B Requests.} A request $r_i$ is served by the local copy in server $s[r_i]$ and this local copy was created later than the preceding request $r_{p(i)}$ (see Figure \ref{12}(b)).
Let $\tau$ be the creation time of
the copy in $s[r_i]$. Then $s[r_i]$ must be receiving a
transfer from another server $s_x$ at time $\tau$.
Since $s[r_i]$ does not have a local request at time $\tau$, by
Proposition \ref{pro-1}, $s_x$ must have a local request $r_k$
at time $\tau$. 

\medskip
\noindent {\bf Type-C Requests.} A request $r_i$ is served by a transfer from another server $s_x$ and the copy at $s_x$ was created no later than the most recent request $r_h$ at $s_x$ before the transfer (see Figure \ref{12}(c)). 
That is, $s_x$ holds a copy from time $t_h$ to $t_i$ and then transfers the object to server $s[r_i]$ to serve $r_i$. 

\medskip
\noindent {\bf Type-D Requests.} A request $r_i$ is served by a transfer from another server $s_x$ and the copy at $s_x$ was created later than the most recent request $r_h$ at $s_x$ before the transfer (see Figure \ref{12}(d)). Let $\tau$ be the creation
time of the copy in $s_x$. Then $s_x$ must be receiving
a transfer from another server $s_y$ at time $\tau$.
Since $s_x$ does not have a local request at time $\tau$, by
Proposition \ref{pro-1}, $s_y$ must have a local request $r_k$
at time $\tau$. 

Since each request must be served by either a transfer or a local copy, it is apparent that the above request categorization is complete.
To prove the induction step, we consider separately the four possible types of request $r_i$ in the optimal offline strategy for the request subsequence $\left\langle r_{1},r_{2},...,r_{i} \right\rangle$. To facilitate presentation, if a data copy is consistently stored in a server before and after a time instant $t$, we say that this copy \textit{crosses} time $t$.

\medskip
\noindent \textbf{Type-A Case}: $r_{i}$ is a \textbf{Type-A} request in the optimal offline strategy, i.e., server $s[r_i]$ holds a copy from the preceding request $r_{p(i)}$ to $r_i$ 
(Figure \ref{12}(a)).

\begin{figure}[htbp]
\centering
\includegraphics[width=8cm]{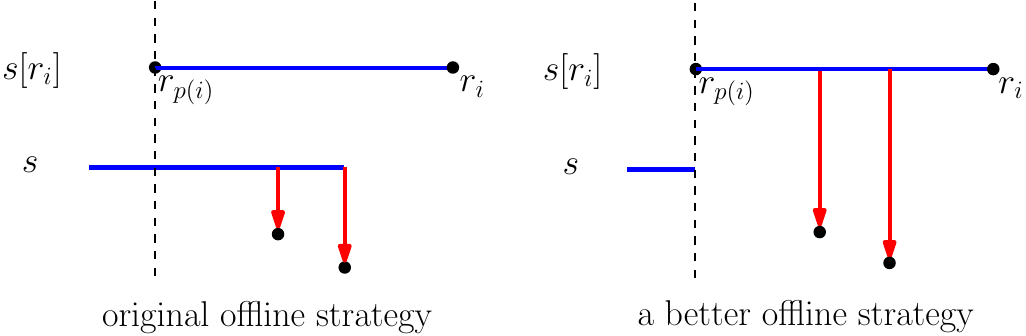}
\caption{Illustration of a data copy crossing time $t_{p(i)}$}
\label{fig_crossing}
\end{figure}

In the optimal offline strategy, if a data copy at some other server $s$ ($s \neq s[r_i]$) crosses time $t_{p(i)}$, the copy must be kept till at least the first local request at $s$ after $t_{p(i)}$ and hence will serve that request. 
In fact, if it does not serve any local request at $s$ (see Figure \ref{fig_crossing} for an illustration), the copy can be deleted earlier without affecting the service of all requests.
This is because all the transfers originating from the copy at $s$ after $t_{p(i)}$ can originate from the copy at $s[r_i]$ instead. 
As a consequence, it contradicts the cost optimality of the offline strategy. Thus, each copy crossing time $t_{p(i)}$ must be kept till at least the first local request after $t_{p(i)}$.

\begin{figure}[htbp]
\centering
\includegraphics[width=4cm]{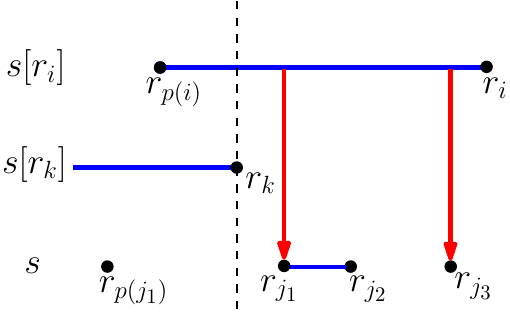}
\caption{$r_i$ is a \textbf{Type-A} request in the optimal offline strategy}
\label{d}
\end{figure}

In the optimal offline strategy, if there is at least one data copy in other servers crossing time $t_{p(i)}$, among all the servers with such data copies, we find the server whose first local request after $t_{p(i)}$ has the highest index (i.e., arises the latest) and denote this request by $r_k$ (where $p(i)<k<i$) and this server by $s[r_k]$ (see Figure \ref{d}).
If there is no data copy in other servers crossing time $t_{p(i)}$, we define $k = p(i)$.
In what follows, we will calculate \textbf{Online$(k+1,i)$} and \textbf{OPT$(k+1,i)$}, and show that $\frac{\textbf{Online$(k+1,i)$}}{\textbf{OPT$(k+1,i)$}} \leq \max\{2, \min\{\gamma, 3\}\}$. Then, together with the induction hypothesis that $\frac{\textbf{Online$(1,k)$}}{\textbf{OPT$(1,k)$}}\leq \max\{2, \min\{\gamma, 3\}\}$, we can conclude that 
$\frac{\textbf{Online$(1,i)$}}{\textbf{OPT$(1,i)$}} 
\leq \max\{2, \min\{\gamma, 3\}\}$. 

We define the set of requests $Q:=\left\{r_{k+1}, r_{k+2},..., r_{i-1}\right\}$, and divide it into $Q_1 \sim Q_6$ based on the request categorization in the online algorithm, i.e., $Q_i$ includes all the \textbf{Type-$i$} requests in $Q$.

We first calculate the offline cost \textbf{OPT$(k+1,i)$}. 
For each request $r_j \in Q$, either (1) $r_j$ is the first local request at server $s[r_j]$ after time $t_{p(i)}$ and by the definition of $k$, no data copy is maintained at $s[r_j]$ crossing time $t_{p(i)}$, or (2) $r_j$ is not the first local request at server $s[r_j]$ after time $t_{p(i)}$. 

In scenario (1), an inward transfer to $s[r_j]$ is required to serve $r_j$ in the optimal offline strategy, which costs $\lambda$. The best way is to transfer at the time of $r_j$ (since an earlier transfer would give rise to unnecessary storage cost at $s[r_j]$). In addition, we must have $t_{j}-t_{p(j)} > \frac{\lambda}{\mu[r_{j}]}$ (otherwise, by Proposition \ref{prop5}, a data copy must be maintained at $s[r_j]$ over the period $(t_{p(j)}, t_{j})$ which crosses time $t_{p(i)}$ in the optimal offline strategy, leading to a contradiction). Thus, it follows from Proposition \ref{pro-3} that $r_j$ is not a \textbf{Type-4} request by the online algorithm. For example, in Figure \ref{d}, $r_{j_1}$ is the first request at $s$ after time $t_{p(i)}$, so it must hold that $t_{j_1}-t_{p(j_1)}>\frac{\lambda}{\mu(s)}$ and $r_{j_1}$ is served by a transfer.  

In scenario (2), if $t_{j}-t_{p(j)}\leq\frac{\lambda}{\mu[r_{j}]}$, by Proposition \ref{prop5}, $r_j$ must be served by the local copy stored in $s[r_j]$ over the period $(t_{p(j)}, t_j)$ in the optimal offline strategy, which costs $\mu[r_{j}]\cdot\big(t_{j}-t_{p(j)}\big)$. In this case, it follows from Proposition \ref{pro-3} that $r_j$ is a \textbf{Type-4} request by the online algorithm. If $t_{j}-t_{p(j)} > \frac{\lambda}{\mu[r_{j}]}$, maintaining a copy in $s[r_{j}]$ over the period $(t_{p(j)}, t_{j})$ has a storage cost more than $\lambda$. Note that a data copy is stored in $s[r_{i}]$ during $(t_{p(i)}, t_{i})$ in the optimal offline strategy. This implies that the best way to serve $r_j$ is by a transfer from $s[r_{i}]$ to $s[r_{j}]$ at the time of $r_j$, which costs $\lambda$. In this case, by Proposition \ref{pro-3}, $r_j$ is not a \textbf{Type-4} request by the online algorithm. For example, in Figure \ref{d}, since $t_{j_2}-t_{j_1}\leq\frac{\lambda}{\mu(s)}$, $r_{j_2}$ must be served locally; since $t_{j_3}-t_{j_2}>\frac{\lambda}{\mu(s)}$, $r_{j_3}$ must be served by a transfer.

\begin{figure}[htbp]
\centering
\includegraphics[width=8cm]{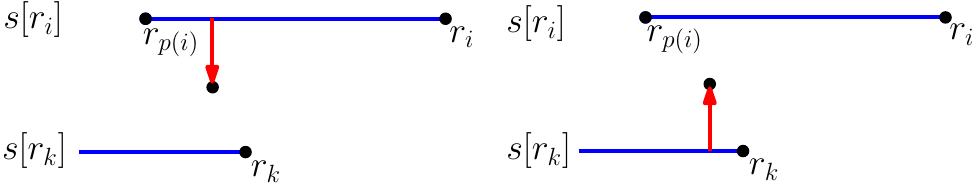}
\caption{\label{equi} Illustration of replacing outward transfers from $s[r_i]$ by those from $s[r_k]$ during the period $(t_{p(i)}, t_k)$}
\end{figure}

Overall, in the optimal offline strategy, the total cost of the data copies and transfers to serve the requests in $Q$ can be written as 
\begin{equation*}
\lambda\cdot 
(|Q_1| + |Q_2| + |Q_3| + |Q_5| + |Q_6|) + \sum_{r_j \in Q_4}
\mu[r_{j}]\cdot\big(t_{j}-t_{p(j)}\big).
\end{equation*}
If we remove these data copies and transfers, all the requests $\langle r_{1}, r_{2},..., r_{k}\rangle$ can still be served. In addition, since there is a data copy in server $s[r_k]$ crossing time $t_{p(i)}$, all outward transfers from server $s[r_i]$ during the period $(t_{p(i)}, t_k)$ can originate from $s[r_k]$ instead with the same cost (see Figure \ref{equi} for an illustration). Hence, we can also remove the data copy in $s[r_i]$ during $(t_{p(i)}, t_{i})$ without affecting the service of the requests $\langle r_{1}, r_{2},..., r_{k}\rangle$. Thus, we have the following relation:
\begin{equation*}
\begin{aligned}
& \textbf{OPT$(1,i)$} - \lambda\cdot 
(|Q_1| + |Q_2| + |Q_3| + |Q_5| + |Q_6|) - \mu[r_{i}]\cdot\big(t_{i}-t_{p(i)}\big) \\ & 
- \sum_{r_j \in Q_4}
\mu[r_{j}]\cdot\big(t_{j}-t_{p(j)}\big) \geq \textbf{OPT$(1,k)$}.
\end{aligned}
\end{equation*}
As a result,
\begin{eqnarray}
 \textbf{OPT$(k+1,i)$} & \geq & \lambda\cdot 
 (|Q_1| + |Q_2| + |Q_3| + |Q_5| + |Q_6|) \nonumber \\& & 
+ \sum_{r_j \in Q_4 \cup\{r_i\}}
\mu[r_{j}]\cdot\big(t_{j}-t_{p(j)}\big). 
\label{1}
\end{eqnarray}

To calculate the online cost \textbf{Online$(k+1,i)$}, 
we check the duration of the period
$(t_{p(i)},t_{i})$. 

(a) If $t_{i}-t_{p(i)}\leq\frac{\lambda}{\mu[r_i]}$, by Proposition \ref{pro-3}, $r_i$ is a \textbf{Type-4} request. In the online algorithm, the data copy stored in $s[r_i]$ during $(t_{p(i)},t_{i})$ is a regular copy. By Proposition \ref{pro-2}, there is no special copy in the system during $(t_{p(i)},t_{i})$ and hence no \textbf{Type-2/3/5/6} request in $Q$. 
Then based on Proposition \ref{costsummary},
the total online cost allocated to the requests in $Q$ and $r_{i}$ is 
\begin{equation*}
\begin{aligned}
& \textbf{Online$(k+1,i)$} = 2\lambda\cdot
|Q_1| 
+ \sum_{r_j \in Q_4 \cup\{r_i\}}
\mu[r_{j}]\cdot\big(t_{j}-t_{p(j)}\big) \\ & \leq^{\text{by (\ref{1})}} 2 \cdot \textbf{OPT$(k+1,i)$}.
\end{aligned}
\end{equation*}

(b) If $t_{i}-t_{p(i)}>\frac{\lambda}{\mu[r_i]}$, based on Proposition \ref{costsummary},
the total online cost allocated to the requests in $Q$ 
is 
\begin{eqnarray}
\lefteqn{\textbf{Online$(k+1,i-1)$}} \nonumber \\ & = & \max\{2, \min\{\gamma, 3\}\}\cdot\lambda \cdot 
(|Q_1| + |Q_2| + |Q_3| + |Q_5| + |Q_6|) \nonumber \\ & & 
+ \!\!\!\!\!\! \sum_{r_j \in Q_2 \cup Q_3 \cup Q_5 \cup Q_6}
\!\!\!\!\!\!\!\!\!\!\!\! \min\{\gamma, 3\} \cdot \mu(s_1)\cdot\big(t_{j}-t'_{j}\big) 
+ \!\!\! \sum_{r_j \in Q_4}
\mu[r_{j}]\cdot\big(t_{j}-t_{p(j)}\big). \nonumber
\end{eqnarray}
Since $t_{i}-t_{p(i)}>\frac{\lambda}{\mu[r_i]}$, by Proposition \ref{pro-3}, $r_{i}$ is not a \textbf{Type-4} request.
Hence, by Proposition \ref{costsummary}, the first term in the allocated cost of $r_{i}$ is bounded by $\max\{2, \min\{\gamma, 3\}\}\cdot\lambda$. 
Therefore, it follows 
that 
\begin{eqnarray}
\lefteqn{\textbf{Online$(k+1, i)$}} \nonumber \\ 
& \leq & \max\{2, \min\{\gamma, 3\}\}\cdot\lambda \cdot 
(|Q_1| + |Q_2| + |Q_3| + |Q_5| + |Q_6|) \nonumber \\ & & +\max\{2, \min\{\gamma, 3\}\}\cdot\lambda 
+ \!\!\!\!\!\! \sum_{r_j \in Q_2 \cup Q_3 \cup Q_5 \cup Q_6 \cup\{r_i\}}
\!\!\!\!\!\!\!\!\!\!\!\!\!\!\!\!\!\! \min\{\gamma, 3\} \cdot \mu(s_1)\cdot\big(t_{j}-t'_{j}\big) \nonumber \\
& & + \sum_{r_j \in Q_4}
\mu[r_{j}]\cdot\big(t_{j}-t_{p(j)}\big). 
\label{2}
\end{eqnarray}

Note that the third and fourth terms above are the storage costs of special copies. We can bound them by using Proposition \ref{pro-2}.

\begin{Proposition} 
It holds that
\begin{equation*}
\sum_{r_j \in Q_2 \cup Q_3 \cup Q_5 \cup Q_6 \cup \{r_i\}}
\big(t_{j}-t'_{j}\big) \leq t_{i}-\Big(t_{p(i)}+\frac{\lambda}{\mu[r_{i}]}\Big).
\end{equation*}
\label{lemma1}
\end{Proposition}
\begin{proof}
In the online algorithm, according to Proposition \ref{pro-2}, the storage periods $(t'_j, t_j)$ of all the special copies relevant to \textbf{Type-2/3/5/6} requests $r_j \in Q\cup\{r_i\}$ do not overlap. Apparently, all these storage periods end before or at the time $t_i$ of $r_i$. Also note that after serving $t_{p(i)}$, there is a regular copy at server $s[r_{p(i)}]$ over the period $\Big(t_{p(i)}, t_{p(i)}+\frac{\lambda}{\mu[r_{i}]}\Big)$ (where $\mu[r_{i}] = \mu[r_{p(i)}]$ because $r_i$ and $r_{p(i)}$ arise at the same server). By Proposition \ref{pro-2} again, the storage period of this regular copy does not overlap with that of any special copy. Thus, the storage periods of all the aforementioned special copies must start no earlier than time $t_{p(i)}+\frac{\lambda}{\mu[r_{i}]}$. 
Hence, the proposition follows.
\end{proof}

It follows from (\ref{2}) and Proposition \ref{lemma1} as well as $\min\{\gamma, 3\} \leq \max\{2, \min\{\gamma, 3\}\}$ and $\mu(s_1) \leq \mu[r_i]$ that
\begin{eqnarray*}
\lefteqn{\textbf{Online$(k+1, i)$}} \nonumber \\
& \leq & \max\{2, \min\{\gamma, 3\}\}\cdot\lambda \cdot 
(|Q_1| + |Q_2| + |Q_3| + |Q_5| + |Q_6|) \nonumber \\ & & +\max\{2, \min\{\gamma, 3\}\}\cdot\lambda \nonumber \\
& & + \max\{2, \min\{\gamma, 3\}\}\cdot\mu[r_{i}]\cdot\bigg(t_{i}-\Big(t_{p(i)}+\frac{\lambda}{\mu[r_{i}]}\Big)\bigg) \nonumber \\
& & + \sum_{r_j \in Q_4}
\mu[r_{j}]\cdot\big(t_{j}-t_{p(j)}\big) \nonumber \\
& = & \max\{2, \min\{\gamma, 3\}\}\cdot\lambda \cdot 
(|Q_1| + |Q_2| + |Q_3| + |Q_5| + |Q_6|) \nonumber \\ & & + \max\{2, \min\{\gamma, 3\}\}\cdot\mu[r_{i}]\cdot\big(t_{i}-t_{p(i)}\big) \nonumber \\
& & + \sum_{r_j \in Q_4}
\mu[r_{j}]\cdot\big(t_{j}-t_{p(j)}\big) \nonumber \\
& \leq^{\text{by (\ref{1})}} & \max\{2, \min\{\gamma, 3\}\} \cdot \textbf{OPT$(k+1,i)$}.
\end{eqnarray*}

\begin{figure}[htbp]
\centering
\includegraphics[width=4cm]{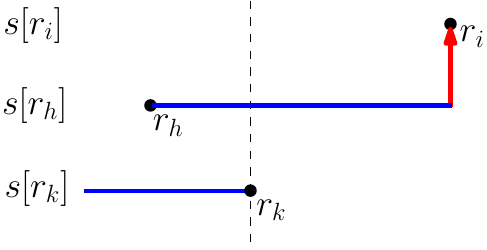}
\caption{$r_i$ is a \textbf{Type-C} request in the optimal offline strategy}
\label{h}
\end{figure}

\medskip
\noindent \textbf{Type-C Case}: If $r_{i}$ is a  \textbf{Type-C} request in the optimal offline strategy (Figure \ref{12}(c)),
the analysis 
is almost the same as the previous \textbf{Type-A} case. 
There can be some data copies crossing time $t_h$. Among such 
copies, we find the server whose first local request after $t_h$ has the highest index 
and denote this request by $r_k$ (see Figure \ref{h} for an illustration). 
Then we can calculate \textbf{Online$(k+1,i)$} and \textbf{OPT$(k+1,i)$}, and show that $\frac{\textbf{Online$(k+1,i)$}}{\textbf{OPT$(k+1,i)$}} \leq \max\{2, \min\{\gamma, 3\}\}$. 
The only difference from the \textbf{Type-A} case is that there is an additional transfer 
at time $t_i$ for serving request $r_i$ in the optimal offline strategy. This gives an additive term of $\lambda$ in \textbf{OPT$(k+1,i)$} which would not affect verifying the bound $\max\{2, \min\{\gamma, 3\}\}$. The complete analysis is provided in Appendix \ref{typec}.

\medskip
\noindent \textbf{Type-B Case}: $r_{i}$ is a \textbf{Type-B} request in the optimal offline strategy, i.e., $r_{i}$ is served by a local copy which is created by a transfer from another server when there is a request $r_{k}$ (where $k < i$) at that server (Figure \ref{12}(b)). 

\begin{figure}[htbp]
\centering
\includegraphics[width=8cm]{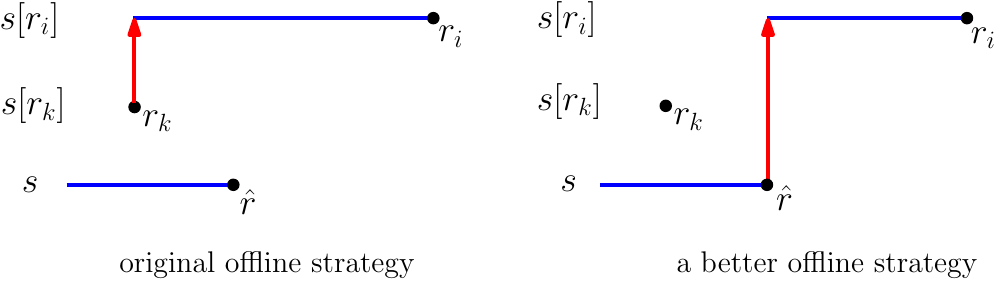}
\caption{Illustration of a data copy crossing time $t_{k}$}
\label{ill}
\end{figure}

In this case, since there is no data copy in server $s[r_i]$ right before time $t_k$, by Proposition \ref{prop5}, we must have $t_{i}-t_{p(i)}>\frac{\lambda}{\mu[r_i]}$ if $r_{p(i)}$ exists. Moreover, it is impossible to have any 
copy that crosses time $t_{k}$. 
This can be proved by contradiction. Assume on the contrary that a copy in some other server $s$ ($s \neq s[r_k]$ and $s \neq s[r_i]$) crosses $t_k$ in the optimal offline strategy. By similar arguments to Figure \ref{fig_crossing}, the copy must be kept till at least 
the first local request $\hat{r}$ at $s$ after $t_{k}$ 
and will serve $\hat{r}$. By letting $s$ transfer the object to server $s[r_i]$ after serving $\hat{r}$, we can save the storage cost in $s[r_i]$ 
(see Figure \ref{ill}). 
This contradicts that the offline strategy is optimal. 

The rest of the analysis is generally similar to the \textbf{Type-A} case.
We will calculate \textbf{Online$(k+1,i)$} and \textbf{OPT$(k+1,i)$}, and show that $\frac{\textbf{Online$(k+1,i)$}}{\textbf{OPT$(k+1,i)$}} \leq \max\{2, \min\{\gamma, 3\}\}$.

Again, let the set of requests  $Q:=\left\{r_{k+1}, r_{k+2},..., r_{i-1}\right\}$, and divide it into $Q_1 \sim Q_6$ based on the request categorization in the online algorithm, i.e., $Q_i$ includes all the \textbf{Type-$i$} requests in $Q$.

By similar arguments to the \textbf{Type-A} case, the total cost of the data copies and transfers to serve the requests in $Q$ in the optimal offline strategy is
\begin{equation*}
\lambda\cdot 
(|Q_1| + |Q_2| + |Q_3| + |Q_5| + |Q_6|) + \sum_{r_j \in Q_4}
\mu[r_{j}]\cdot\big(t_{j}-t_{p(j)}\big).
\end{equation*}

If we remove these data copies and transfers as well as the copy in $s[r_i]$ during $(t_k, t_i)$ and the transfer from $s[r_k]$ to $s[r_i]$, all the requests $\langle r_{1}, r_{2},..., r_{k}\rangle$ can still be served. Thus,
\begin{equation}
\begin{aligned}
& \textbf{OPT$(k+1,i)$} \geq \lambda\cdot 
(|Q_1| + |Q_2| + |Q_3| + |Q_5| + |Q_6|) +\lambda \\ & + \mu[r_{i}]\cdot\big(t_{i}-t_{k}\big) + \sum_{r_j \in Q_4}
\mu[r_{j}]\cdot\left(t_{j}-t_{p(j)}\right).
\label{6}
\end{aligned}
\end{equation}

Since $t_{i}-t_{p(i)}>\frac{\lambda}{\mu[r_i]}$, based on Proposition \ref{costsummary}, 
\textbf{Online$(k+1,i)$} has the same bound as given in (\ref{2}). 

\begin{Proposition} 
It holds that
\begin{equation*}
\sum_{r_j \in Q_2 \cup Q_3 \cup Q_5 \cup Q_6 \cup \{r_i\}}
\big(t_{j}-t'_{j}\big) \leq t_{i}-t_{k}.
\end{equation*}
\label{lemma2}
\end{Proposition}
\begin{proof}
The proof is similar to Proposition \ref{lemma1}. 
\end{proof}

It follows from (\ref{2}) and Proposition \ref{lemma2} as well as $\mu(s_1) \leq \mu[r_i]$ that
\begin{eqnarray*}
\lefteqn{\textbf{Online$(k+1, i)$}} \nonumber \\ 
& \leq & \max\{2, \min\{\gamma, 3\}\}\cdot\lambda \cdot 
(|Q_1| + |Q_2| + |Q_3| + |Q_5| + |Q_6|) \nonumber \\ 
& & + \max\{2, \min\{\gamma, 3\}\}\cdot\lambda 
+ \min\{\gamma, 3\}\cdot\mu[r_{i}]\cdot\left(t_{i}-t_{k}\right) \nonumber \\
& & + \sum_{r_j \in Q_4}
\mu[r_{j}]\cdot\big(t_{j}-t_{p(j)}\big) \nonumber \\
& \leq^{\text{by (\ref{6})}} & \max\{2, \min\{\gamma, 3\}\} \cdot \textbf{OPT$(k+1,i)$}. 
\end{eqnarray*}

\medskip
\noindent \textbf{Type-D Case}: If $r_{i}$ is a \textbf{Type-D} request in the optimal offline strategy (Figure \ref{12}(d)),
the analysis 
is almost the same as the previous \textbf{Type-B} case. 
There is no data copy crossing time $t_k$. 
The only difference from the \textbf{Type-B} case is that there is an additional transfer %of cost $\lambda$ 
at time $t_i$ for serving request $r_i$ in the optimal offline strategy. This gives an additive term of $\lambda$ in \textbf{OPT$(k+1,i)$} which would not affect verifying the bound $\frac{\textbf{Online$(k+1,i)$}}{\textbf{OPT$(k+1,i)$}} \leq \max\{2, \min\{\gamma, 3\}\}$. The complete analysis is provided in Appendix \ref{typed}.

To conclude, 
Algorithm \ref{alg3} 
is $\max\{2, \min\{\gamma, 3\}\}$-competitive. 

\subsubsection{Tight Examples}

We give some examples to show that the competitive analysis is tight. 
Consider two servers $s_1$ and $s_2$. 

\begin{figure}[htbp]
\centering
\includegraphics[width=6cm]{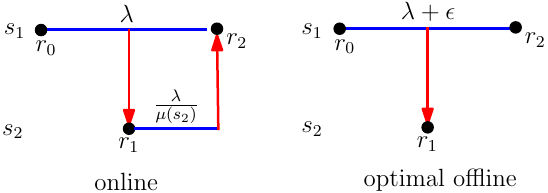}
\caption{An example when $\gamma \leq 2$}
\label{ill1}
\end{figure}

In the first example (Figure \ref{ill1}), $\mu(s_1)=1$ and $1 < \mu(s_2) \leq 2$ so that $\gamma \leq 2$. The initial copy is in $s_1$. Two requests $r_1$ and $r_2$ arise at $s_2$ and $s_1$ respectively at times $t_1 = \big(1-\frac{1}{\mu(s_2)}\big)\cdot\lambda+\epsilon$, $t_2 = \lambda+\epsilon$ where $\epsilon > 0$. By our online algorithm, the regular copy in $s_1$ expires before request $r_2$, 
so both $r_1$ and $r_2$ are served by transfers. In the optimal offline strategy, $s_1$ should keep the copy till $r_2$ and serve $r_1$ by a transfer. Hence, the online-to-optimal cost ratio is $\frac{4\lambda}{2\lambda+\epsilon}\rightarrow 2$ as $\epsilon\rightarrow 0$.

\begin{figure}[htbp]
\centering
\includegraphics[width=8cm]{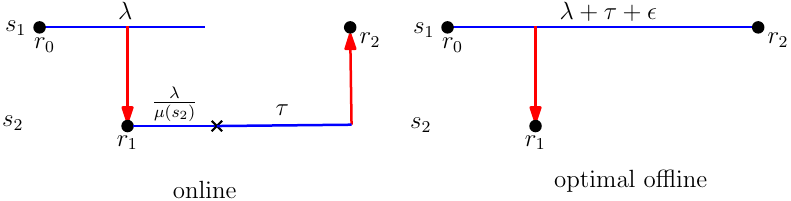}
\caption{An example when $2 < \gamma \leq 3$}
\label{ill2}
\end{figure}

In the second example (Figure \ref{ill2}), $\mu(s_1)=1$ and $2 < \mu(s_2) \leq 3$ so that $2< \gamma \leq 3$. The initial copy is in $s_1$. Two requests $r_1$ and $r_2$ arise at $s_2$ and $s_1$ respectively at times $t_1 = \big(1-\frac{1}{\mu(s_2)}\big)\cdot\lambda+\epsilon$, $t_2 = \lambda+\tau+\epsilon$ where $\tau, \epsilon > 0$. By our online algorithm, the regular copy in $s_2$ expires after that of $s_1$, so $s_2$ keeps a resident special copy till request $r_2$, and serves it by a transfer. In the optimal offline strategy, $s_1$ should keep the copy till $r_2$ and serve $r_1$ by a transfer. Hence, the online-to-optimal cost ratio is $\frac{4\lambda+\mu(s_2)\cdot \tau}{2\lambda+t'+\epsilon}\rightarrow \mu(s_2)=\gamma$ as $\tau \rightarrow \infty$ and $\epsilon \rightarrow 0$.

\begin{figure}[htbp]
\centering
\includegraphics[width=5cm]{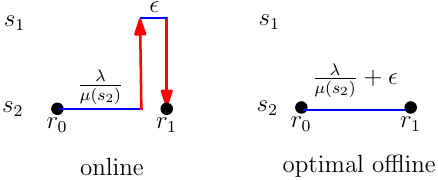}
\caption{An example when $ \gamma > 3$}
\label{ill3}
\end{figure}

In the third example (Figure \ref{ill3}), $\mu(s_1)=1$ and $\mu(s_2) > 3$ so that $\gamma > 3$. The initial copy is in $s_2$. One request $r_1$ arises at $s_2$ at time $t_1 = \frac{\lambda}{\mu(s_2)}+\epsilon$ where $\epsilon > 0$. By our online algorithm, the regular copy in $s_2$ expires before request $r_1$. Since $\mu(s_2)>3$, $s_2$ transfers the object to $s_1$ to create a relocated special copy which is kept till $r_1$ and serves it by another transfer. In the optimal offline strategy, $s_2$ should keep the copy till $r_1$. Hence, the online-to-optimal cost ratio is $\frac{3\lambda+\epsilon}{\lambda+\mu(s_2)\cdot\epsilon}\rightarrow 3$ as $\epsilon \rightarrow 0$.

\section{Experimental Evaluation}
\label{experiment}
To evaluate the empirical performance of Algorithm~\ref{alg3}, 
we conduct simulation experiments using object access traces from a cloud-based storage service provided by IBM~\cite{IBM}. As similar performance trends are observed across different traces in the experiments, we present the results for a representative trace of an object with ID ``cdb1824b71efe7d4'' in a dataset ``IBM Object Store Trace Number 003''. 
This object received a total of 11,683 read requests over a period of 7 days in 2019. We rescale the time axis by treating one second as a time unit, resulting in a total simulation duration of 604,800 time units.

At the beginning of the simulation, only one copy of the object is placed at server $s_1$. The requests are uniformly distributed at random across 10 servers. Under this distribution, 
the average inter-request time per server is estimated to be around 500 time units.

We compare Algorithm~\ref{alg3} with the online algorithm proposed by Wang~\textit{et al.}~\cite{2023Cost, 2021Cost}. 
Additionally, we also include a simple benchmark algorithm (referred to as the \textit{simple algorithm} hereafter), which always keeps a data copy at server $s_1$, and stores a regular copy for a duration of $\lambda / \mu(s_j)$ after each local request at any other server $s_j \neq s_1$. It can be shown that this algorithm is 3-competitive.\footnote{In the worst case, there is no request at $s_1$, and all requests at other servers are served by transfer in the simple algorithm. The total storage and transfer cost on all servers except $s_1$ is at most $2 \cdot \textbf{OPT}$, while the permanent copy at $s_1$ incurs an additional storage cost of at most $\textbf{OPT}$, yielding a total cost of $3 \cdot \textbf{OPT}$.}

We evaluate the algorithms under different settings of storage cost rates of the 10 servers:
\begin{itemize}
\item \textbf{Set 1:} \{1, 1, 1, 1, 1, 1, 1, 1, 1, 1\} (all rates are equal),
\item \textbf{Set 2:} \{1, 1.1, 1.2, 1.3, 1.3, 1.4, 1.5, 1.7, 2.1, 2.3\} (all rates are bounded by 3),
\item \textbf{Set 3:} \{1, 1.1, 1.2, 1.5, 1.6, 2.1, 2.3, 2.7, 3.1, 4\} (a mix of rates in different ranges),
\item \textbf{Set 4:} \{1, 1.1, 1.2, 1.3, 1.5, 2.1, 3, 6, 10, 15\} (a mix of rates in different ranges),
\end{itemize}
and different transfer costs \{50, 75, ..., 1175, 1200\} (which cover a wide range of values below and above the average inter-request time at each server). We normalize the online cost incurred by each algorithm against the cost of the optimal offline strategy~\cite{2021Cost}, and call it the \textit{online-to-optimal ratio}.

\begin{figure}[htbp]
    \centering
    \begin{minipage}[t]{0.46\textwidth}
        \centering
        \includegraphics[width=1\textwidth]{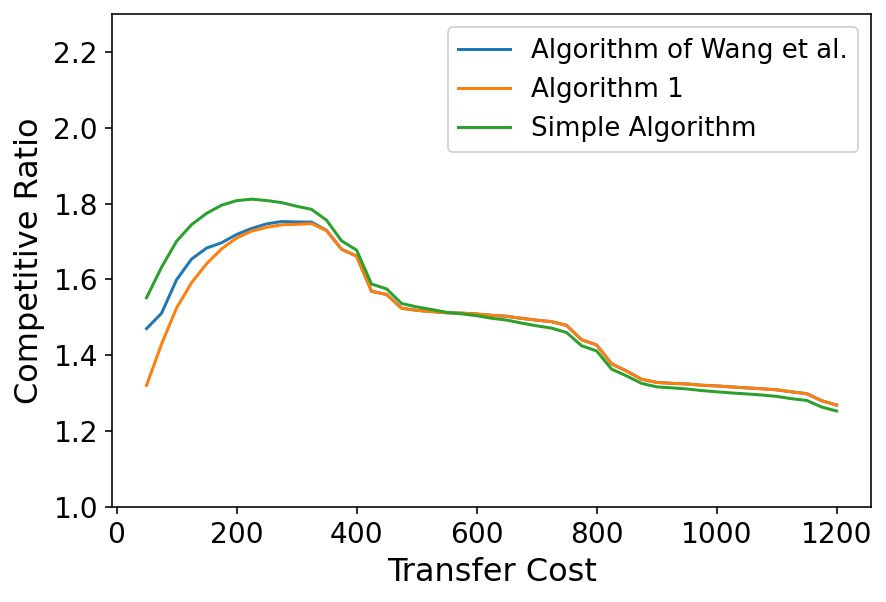}
        \caption{Results of Set 1 of storage cost rates}
        \label{set1}
    \end{minipage}
    \hfill
    \begin{minipage}[t]{0.46\textwidth}
        \centering
        \includegraphics[width=1\textwidth]{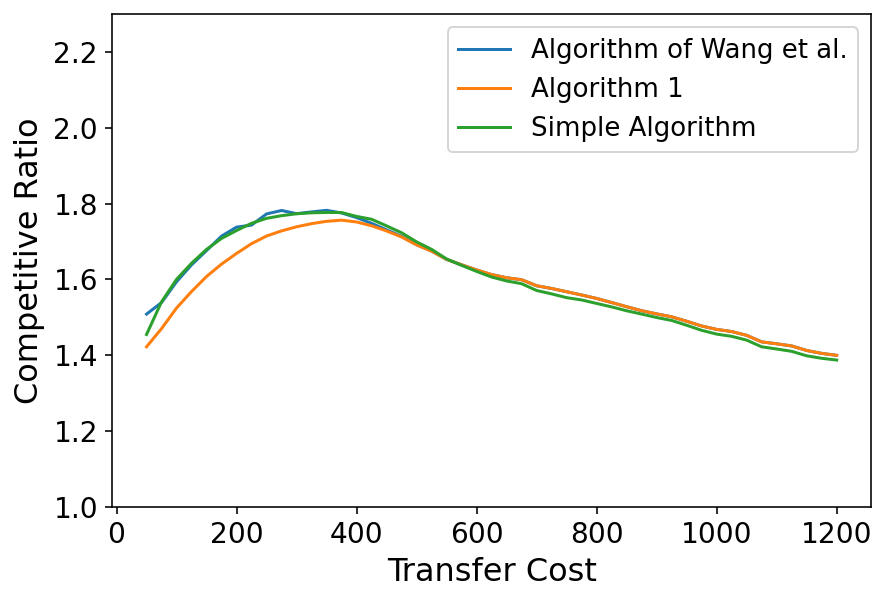}
        \caption{Results of Set 2 of storage cost rates}
        \label{set5}
    \end{minipage}

    \begin{minipage}[t]{0.46\textwidth}
        \centering
        \includegraphics[width=1\textwidth]{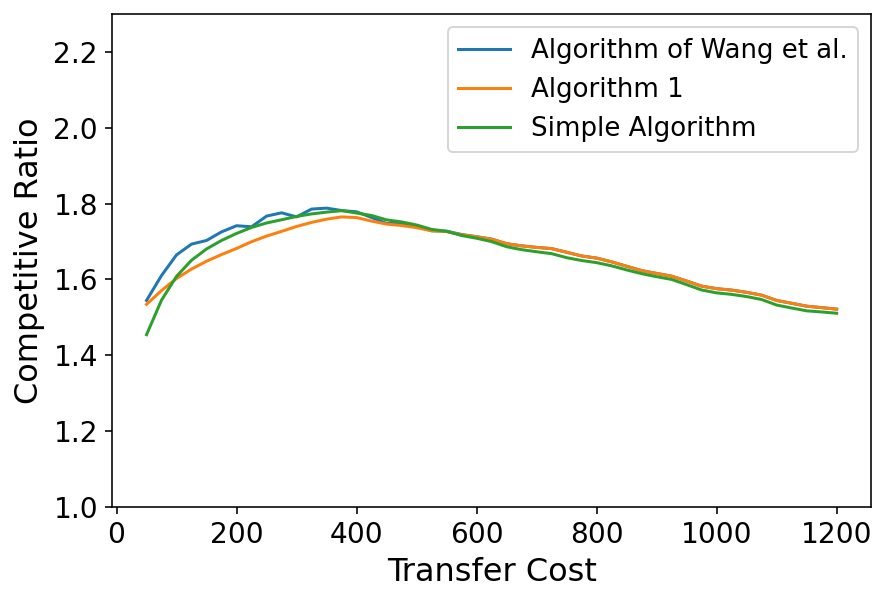}
        \caption{Results of Set 3 of storage cost rates}
        \label{set6}
    \end{minipage}
    \hfill
    \begin{minipage}[t]{0.46\textwidth}
        \centering
        \includegraphics[width=1\textwidth]{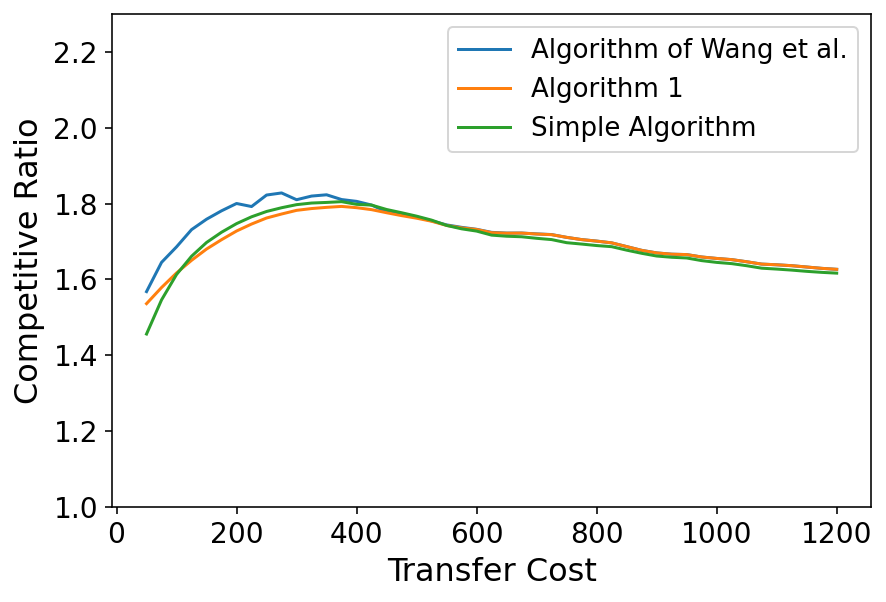}
        \caption{Results of Set 4 of storage cost rates}
        \label{set4}
    \end{minipage}
\end{figure}

Figures~\ref{set1}-\ref{set4} present 
the online-to-optimal ratio under various settings of storage cost rates and transfer costs. 

A common trend observed from Figures \ref{set1}-\ref{set4} is that the 
performance of all three algorithms gradually converges as the transfer cost increases. 
This is because when the transfer cost $\lambda$ exceeds the average inter-request time at server $s_1$ (approximately 500 time units), regular copies at $s_1$ tend to persist long enough to 
cover the next requests.
As a result, 
both Algorithm~\ref{alg3} and Wang~\textit{et al.}'s algorithm~\cite{2023Cost, 2021Cost} imitate 
the simple algorithm which maintains a permanent copy at $s_1$.
The three algorithms also behave similarly at the other servers: requests with short inter-arrival times are served by regular copies, while those with long inter-arrival times are served via transfers. 

For Set 1 and Set 2 where all storage cost rates are less than 3 (see Figures~\ref{set1} and \ref{set5}), when the transfer cost $\lambda$ is below 500,
our proposed Algorithm~\ref{alg3} consistently outperforms Wang~\textit{et al.}'s algorithm and the simple algorithm. 
This is because when the time interval between two successive requests is long, Algorithm~\ref{alg3} keeps a special copy (either resident or relocated) and drops it upon serving the second request. In contrast, Wang~\textit{et al.}'s algorithm retains the copy for some fixed duration and does not drop it after serving the second request. 
This results in redundant storage cost, since a new regular copy will also be created after the second request. 
Similarly, the simple algorithm always keeps a copy at server $s_1$, which also incurs unnecessary storage cost when $s_1$ is not frequently accessed. Therefore, both Wang~\textit{et al.}'s algorithm and the simple algorithm have higher online costs than Algorithm~\ref{alg3}. 

For Set 3 and Set 4 where some servers have storage cost rates above 3 (see Figures~\ref{set6} and \ref{set4}),
our proposed Algorithm~\ref{alg3} again produces lower online-to-optimal ratios than Wang~\textit{et al.}'s algorithm and the simple algorithm in most cases when $\lambda < 500$. 
One interesting trend is that the simple algorithm achieves better relative performance compared to Set 1 and Set 2. This is because if there is only one regular copy in the system at a server with storage cost rate above 3, when it expires, both Algorithm~\ref{alg3} and Wang~\textit{et al.}'s algorithm keep the copy at the same server for some time before transferring it to server $s_1$. Such transfers become more frequent when the regular copy durations are much shorter than inter-request intervals ($\lambda$ is very small). In contrast, the simple algorithm saves these transfer costs by maintaining a permanent copy at $s_1$. 
This observation indicates a potential direction for refining Algorithm~\ref{alg3} to enhance its performance under such conditions.

\section{Concluding Remarks}
In this paper, we have presented an online algorithm for a cost optimization problem of distributed 
data access. 
The algorithm has a competitive ratio of $2$ if the max/min storage cost ratio $\gamma \leq 2$, a competitive ratio of $\gamma$ if $2 < \gamma \leq 3$, and a competitive ratio of $3$ if $\gamma > 3$. 
In addition, we have shown that no deterministic online algorithm can achieve a competitive ratio bounded by $2$ if $\gamma > 4$. An open question for future work is to close the gap between the lower bound and upper bound. 

\section*{Acknowledgments}
This research is supported by the Ministry of Education, Singapore, under its Academic Research Fund Tier 2 (Award MOE-T2EP20122-0007) and Academic Research Fund Tier 1 (Award RG23/23).

\newpage
\appendix

\section{Considering An Infinite Time Horizon}
\label{infinite}
It would not affect the correctness of our competitive analysis even if we insist in considering an infinite time horizon. An optimal offline strategy must eventually leave a copy infinitely at a server $s_k$ that is either $s_1$ (the server with the lowest storage cost rate) or another server having the same storage cost rate as $s_1$. Suppose by our online algorithm, the regular copy expiring the latest is at server $s_h$. If $\mu(s_h)\leq 3\cdot\mu(s_1)$, the online algorithm will keep the copy at $s_h$ infinitely. Since the optimal offline strategy keeps the copy at $s_k$ infinitely, the ratio between the storage costs incurred is bounded by $\min\{\gamma,3\}$. 
If $\mu(s_h)>3\cdot\mu(s_1)$, the online algorithm will move the copy to $s_1$ when the regular copy expires. Since $s_k$ has the same storage cost rate as $s_1$, we can assume the copy is moved to $s_k$ (it does not affect the total cost). Then, both offline and online strategies eventually keep the copy at $s_k$ infinitely. We can add a dummy request at $s_k$ when the copy is at $s_k$ in both strategies (this would not change their costs). The same competitive analysis presented in Section \ref{induction} applies to the request sequence with the dummy request appended. Thus, the competitive ratio is still bounded by $\max\{2,\min\{\gamma,3\}\}$.

\section{Proof of Proposition \ref{pro-1}}
\label{proofreqatsrcdes}
\begin{figure}[htbp]
\centering
\includegraphics[width=5.5cm]{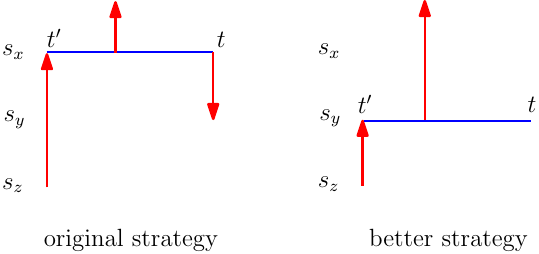}
\caption{\label{eg1} The copy at $s_x$ must serve at least one local request}
\end{figure}

\begin{figure}[htbp]
\centering
\includegraphics[width=5.5cm]{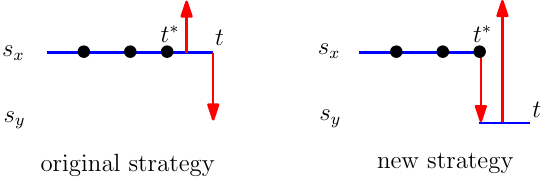}
\caption{\label{eg2} Change the strategy to have a request at the source server of the transfer}
\end{figure}

\begin{proof}

Since one data copy is initially placed in server $s_g$ 
where the dummy request $r_0$ arises at time $0$, no transfer is needed to serve $r_0$.
Thus, we consider only transfers carried out after time $0$.

Suppose that a transfer from a server $s_x$ to a server $s_y$ is
carried out at time $t$ in an optimal offline strategy satisfying Proposition \ref{prop5}, and
there is no request at servers $s_x$ and $s_y$ at time $t$.
It can be inferred that either (1) a data copy is held by $s_x$ before
the transfer, or (2) $s_x$ just receives the data object from
another server $s_z$ at time $t$. In the latter case, since there is
no request at $s_x$ at time $t$, the transfer from $s_x$ to $s_y$
can be replaced by a transfer from $s_z$ to $s_y$ without affecting
the total cost (see Figure \ref{trans1} for an illustration). If $s_z$ does not have a local
request at time $t$ and does not hold a copy before time $t$, we
can find another server transferring to $s_z$ at time $t$ and repeat
the replacement until a source server of the transfer having a local
request at time $t$ or holding a copy before time $t$ is found.
So without loss of generality, we  assume that 
$s_x$ holds a copy before the transfer.

\begin{figure}[htbp]
\centering
\includegraphics[width=4.8cm]{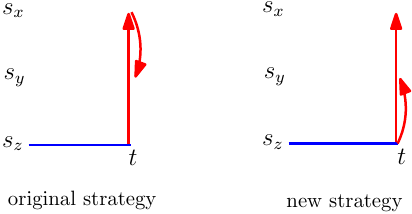}
\caption{\label{trans1} Source server holds a copy before the transfer}
\end{figure}

Similarly, since there is no request at $s_y$ at time $t$, if $s_y$
does not hold a copy after the transfer, $s_y$ must be sending
the data object to another server $s_z$ at time $t$ (otherwise, the
transfer from $s_x$ to $s_y$ can be removed, which contradicts the cost
optimality). We can then replace the
transfers from $s_x$ to $s_y$ and from $s_y$ to $s_z$ by a single
transfer from $s_x$ to $s_z$ without affecting the service of all the requests, which again
contradicts the cost optimality (see Figure \ref{trans2} for an illustration). Thus, $s_y$ must hold a copy
after the transfer.

\begin{figure}[htbp]
\centering
\includegraphics[width=5cm]{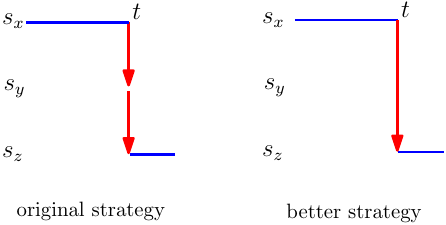}
\caption{\label{trans2} Destination server holds a copy after the transfer}
\end{figure}

If $s_x$ continues to hold a copy after the transfer, we can
delay the transfer from $s_x$ to $s_y$ by some small $\varepsilon >
0$ and create a copy at $s_y$ at time $t+\varepsilon$ to save
the storage cost. This would not affect the feasibility of the
strategy since all the transfers originating from $s_y$
during the period $(t, t+\varepsilon)$ can originate from $s_x$
instead. As a result, it contradicts the cost optimality of the
strategy.

Now suppose that $s_x$ does not hold a copy after the transfer.
If the storage cost rates $u[s_x] > u[s_y]$, we can bring the
transfer from $s_x$ to $s_y$ earlier to time $t - \varepsilon$,
remove the copy at $s_x$ and add a copy at $s_y$ during the
period $(t-\varepsilon, t)$ to save cost.
Similarly, if $u[s_x] < u[s_y]$, we can delay the transfer to
time $t + \varepsilon$, extend the copy at $s_x$ and remove the
copy at $s_y$ during $(t, t+\varepsilon)$ to save cost.
In both cases, it contradicts the cost optimality of the
strategy.

Finally, consider the case $u[s_x] = u[s_y]$.
It can be shown that the copy held by $s_x$ before the transfer serves at least one local request at $s_x$. Otherwise, if the copy held by $s_x$ before the transfer
does not serve any local request at $s_x$, let $t^\prime$ denote the creation
time of the copy. At time $t^\prime$, there must be a transfer
from another server $s_z$ to $s_x$. We can replace the two transfers from $s_z$ to $s_x$ at time $t^\prime$ and
from $s_x$ to $s_y$ at time $t$ by one transfer from $s_z$ to $s_y$ at time
$t^\prime$, remove the copy at $s_x$ and add a copy at $s_y$
during $(t^\prime, t)$, which reduces the total cost, contradicting the cost optimality of the strategy (see Figure \ref{eg1} for an illustration).
Thus, the copy held by $s_x$ before the transfer serves at least one local request at $s_x$.
We look for the last such local request of $s_x$.
Let $t^*$ denote the time of this request. Then we can
bring the transfer from $s_x$ to $s_y$ earlier to time $t^*$,
remove the copy at $s_x$ and add a copy at $s_y$ during the
period $(t^*, t)$ (see Figure \ref{eg2} for an illustration). This would not increase the total cost, and would not affect the service of other requests because all transfers originating from $s_x$ during $(t^*, t)$ can originate from $s_y$ instead.
In the new strategy, there is a request at $s_x$ at the time of the transfer from $s_x$ to $s_y$.
\end{proof}

\section{\textbf{Type-C} Case}
\label{typec}

$r_{i}$ is a \textbf{Type-C} request in the optimal offline strategy, i.e., $r_i$ is served by a transfer from another server $s[r_h]$ which stores a data copy since its most recent request $r_h$ (where $h < i$) (Figure \ref{12}(c)). 

\begin{figure}[htbp]
\centering
\includegraphics[width=8cm]{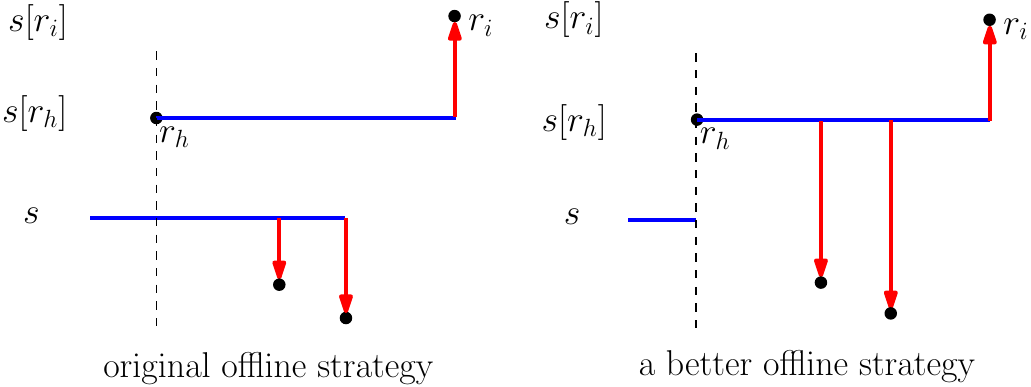}
\caption{Illustration of a data copy crossing time $t_{h}$}
\label{fig_crossingC}
\end{figure}

In the optimal offline strategy, if a data copy at some other server $s$ ($s \neq s[r_h]$) crosses time $t_{h}$, the copy must be kept till at least the first local request at $s$ after $t_{h}$ and hence will serve that request. 
In fact, if it does not serve any local request at $s$ (see Figure \ref{fig_crossingC} for an illustration), the copy can be deleted earlier without affecting the service of all requests.
This is because all the transfers originating from the copy at $s$ after $t_{h}$ can originate from the copy at $s[r_h]$ instead. 
As a consequence, it contradicts the cost optimality of the offline strategy. Thus, each copy crossing time $t_{h}$ must be kept till at least the first local request after $t_{h}$.

\begin{figure}[htbp]
\centering
\includegraphics[width=4cm]{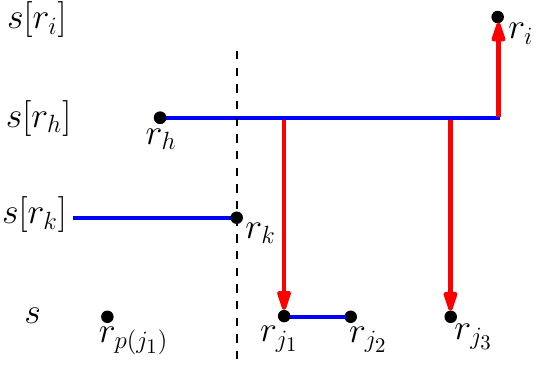}
\caption{$r_i$ is a \textbf{Type-C} request in the optimal offline strategy}
\label{dC}
\end{figure}

In the optimal offline strategy, if there is at least one data copy in other servers crossing time $t_{h}$, among all the servers with such data copies, we find the server whose first local request after $t_{h}$ has the highest index (i.e., arises the latest) and denote this request by $r_k$ (where $h<k<i$) and this server by $s[r_k]$ (see Figure \ref{dC}).
If there is no data copy in other servers crossing time $t_{h}$, we define $k = h$.
In what follows, we will calculate \textbf{Online$(k+1,i)$} and \textbf{OPT$(k+1,i)$}, and show that $\frac{\textbf{Online$(k+1,i)$}}{\textbf{OPT$(k+1,i)$}} \leq \max\{2, \min\{\gamma, 3\}\}$. Then, together with the induction hypothesis that $\frac{\textbf{Online$(1,k)$}}{\textbf{OPT$(1,k)$}}\leq \max\{2, \min\{\gamma, 3\}\}$, we can conclude that 
$\frac{\textbf{Online$(1,i)$}}{\textbf{OPT$(1,i)$}} 
\leq \max\{2, \min\{\gamma, 3\}\}$. 

We define the set of requests $Q:=\left\{r_{k+1}, r_{k+2},..., r_{i-1}\right\}$, and divide it into $Q_1 \sim Q_6$ based on the request categorization in the online algorithm, i.e., $Q_i$ includes all the \textbf{Type-$i$} requests in $Q$.

We first calculate the offline cost \textbf{OPT$(k+1,i)$}. 
For each request $r_j \in Q$, either (1) $r_j$ is the first local request at server $s[r_j]$ after time $t_{h}$ and by the definition of $k$, no data copy is maintained at $s[r_j]$ crossing time $t_{h}$, or (2) $r_j$ is not the first local request at server $s[r_j]$ after time $t_{h}$. 

In scenario (1), an inward transfer to $s[r_j]$ is required to serve $r_j$ in the optimal offline strategy, which costs $\lambda$. The best way is to transfer at the time of $r_j$ (since an earlier transfer would give rise to unnecessary storage cost at $s[r_j]$). In addition, we must have $t_{j}-t_{p(j)} > \frac{\lambda}{\mu[r_{j}]}$ (otherwise, by Proposition \ref{prop5}, a data copy must be maintained at $s[r_j]$ over the period $(t_{p(j)}, t_{j})$ which crosses time $t_{h}$ in the optimal offline strategy, leading to a contradiction). Thus, it follows from Proposition \ref{pro-3} that $r_j$ is not a \textbf{Type-4} request by the online algorithm. For example, in Figure \ref{dC}, $r_{j_1}$ is the first request at $s$ after time $t_h$, so it must hold that $t_{j_1}-t_{p(j_1)}>\frac{\lambda}{\mu(s)}$ and $r_{j_1}$ is served by a transfer.

In scenario (2), if $t_{j}-t_{p(j)}\leq\frac{\lambda}{\mu[r_{j}]}$, by Proposition \ref{prop5}, $r_j$ must be served by the local copy stored in $s[r_j]$ over the period $(t_{p(j)}, t_j)$ in the optimal offline strategy, which costs $\mu[r_{j}]\cdot\big(t_{j}-t_{p(j)}\big)$. In this case, it follows from Proposition \ref{pro-3} that $r_j$ is a \textbf{Type-4} request by the online algorithm. If $t_{j}-t_{p(j)} > \frac{\lambda}{\mu[r_{j}]}$, maintaining a copy in $s[r_{j}]$ over the period $(t_{p(j)}, t_{j})$ has a storage cost more than $\lambda$. Note that a data copy is stored in $s[r_{h}]$ during $(t_{h}, t_{i})$ in the optimal offline strategy. This implies that the best way to serve $r_j$ is by a transfer from $s[r_{h}]$ to $s[r_{j}]$ at the time of $r_j$, which costs $\lambda$. In this case, by Proposition \ref{pro-3}, $r_j$ is not a \textbf{Type-4} request by the online algorithm. For example, in Figure \ref{dC}, since $t_{j_2}-t_{j_1}\leq\frac{\lambda}{\mu(s)}$, $r_{j_2}$ must be served locally; since $t_{j_3}-t_{j_2}>\frac{\lambda}{\mu(s)}$, $r_{j_3}$ must be served by a transfer.

\begin{figure}[htbp]
\centering
\includegraphics[width=8cm]{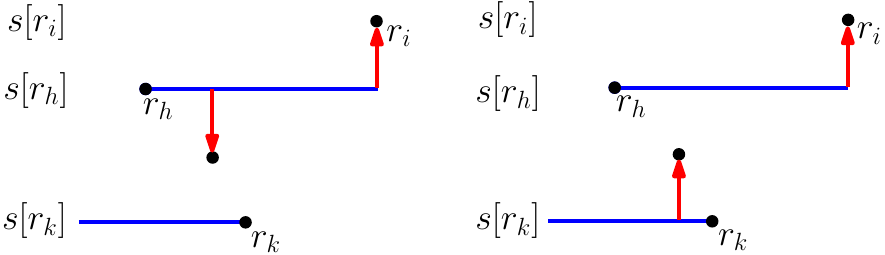}
\caption{\label{equiC} Illustration of replacing outward transfers from $s[r_h]$ by those from $s[r_k]$ during the period $(t_{h}, t_k)$}
\end{figure}

Overall, in the optimal offline strategy, the total cost of the data copies and transfers to serve the requests in $Q$ can be written as 
\begin{equation*}
\lambda\cdot 
(|Q_1| + |Q_2| + |Q_3| + |Q_5| + |Q_6|) + \sum_{r_j \in Q_4}
\mu[r_{j}]\cdot\big(t_{j}-t_{p(j)}\big).
\end{equation*}
If we remove these data copies and transfers, all the requests $\langle r_{1}, r_{2},..., r_{k}\rangle$ can still be served. In addition, since there is a data copy in server $s[r_k]$ crossing time $t_{h}$, all outward transfers from server $s[r_h]$ during the period $(t_{h}, t_k)$ can originate from $s[r_k]$ instead with the same cost (see Figure \ref{equiC} for an illustration). Hence, we can also remove the data copy in $s[r_h]$ during $(t_{h}, t_{i})$ and the transfer from $s[r_h]$ to $s[r_i]$ without affecting the service of the requests $\langle r_{1}, r_{2},..., r_{k}\rangle$. Thus, we have the following relation:
\begin{equation*}
\begin{aligned}
& \textbf{OPT$(1,i)$} - \lambda\cdot 
(|Q_1| + |Q_2| + |Q_3| + |Q_5| + |Q_6|) - \lambda - \mu[r_{h}]\cdot\big(t_{i}-t_{h}\big) \\ & 
- \sum_{r_j \in Q_4}
\mu[r_{j}]\cdot\big(t_{j}-t_{p(j)}\big) \geq \textbf{OPT$(1,k)$}.
\end{aligned}
\end{equation*}
As a result,
\begin{equation}
\begin{aligned}
 &\textbf{OPT$(k+1,i)$} \geq \lambda\cdot 
 (|Q_1| + |Q_2| + |Q_3| + |Q_5| + |Q_6|) + \lambda \\&  
+ \mu[r_{h}]\cdot\big(t_{i}-t_{h}\big) + \sum_{r_j \in Q_4}
\mu[r_{j}]\cdot\big(t_{j}-t_{p(j)}\big).
\label{1C}
\end{aligned}
\end{equation}

To calculate the online cost \textbf{Online$(k+1,i)$}, 
we check the duration of the period
$(t_{h},t_{i})$. 

(a) If $t_{i}-t_{h}\leq\frac{\lambda}{\mu[r_h]}$, 
in the online algorithm, the data copy stored in $s[r_h]$ during $(t_{h},t_{i})$ is a regular copy. By Proposition \ref{pro-2}, there is no special copy in the system during $(t_{h},t_{i})$ and hence no \textbf{Type-2/3/5/6} request in $Q$. Likewise, $r_i$ cannot be a \textbf{Type-2/3/5/6} request. Since $r_i$ is served by a transfer in the optimal offline strategy, by Proposition \ref{prop5}, it must hold that $t_{i}-t_{p(i)}>\frac{\lambda}{\mu[r_i]}$. By Proposition \ref{pro-3}, $r_{i}$ is not a \textbf{Type-4} request. Hence, $r_{i}$ must be a \textbf{Type-1} request.
Then based on Proposition \ref{costsummary},
the total online cost allocated to the requests in $Q$ and $r_{i}$ is 
\begin{equation*}
\begin{aligned}
& \textbf{Online$(k+1,i)$} = 2\lambda\cdot
|Q_1| + 2\lambda 
+ \sum_{r_j \in Q_4}
\mu[r_{j}]\cdot\big(t_{j}-t_{p(j)}\big) \\ & \leq^{\text{by (\ref{1C})}} 2 \cdot \textbf{OPT$(k+1,i)$}.
\end{aligned}
\end{equation*}

(b) If $t_{i}-t_{h}>\frac{\lambda}{\mu[r_h]}$, based on Proposition \ref{costsummary},
the total online cost allocated to the requests in $Q$ 
is 
\begin{eqnarray}
\lefteqn{\textbf{Online$(k+1,i-1)$}} \nonumber \\ & = & \max\{2, \min\{\gamma, 3\}\}\cdot\lambda \cdot 
(|Q_1| + |Q_2| + |Q_3| + |Q_5| + |Q_6|) \nonumber \\ & & 
+ \!\!\!\!\!\! \sum_{r_j \in Q_2 \cup Q_3 \cup Q_5 \cup Q_6}
\!\!\!\!\!\!\!\!\!\!\!\! \min\{\gamma, 3\} \cdot \mu(s_1)\cdot\big(t_{j}-t'_{j}\big) 
+ \sum_{r_j \in Q_4}
\!\!\! \mu[r_{j}]\cdot\big(t_{j}-t_{p(j)}\big). \nonumber
\end{eqnarray}
Since $r_i$ is served by a transfer in the optimal offline strategy, by Proposition \ref{prop5}, it must hold that $t_{i}-t_{p(i)}>\frac{\lambda}{\mu[r_i]}$. By Proposition \ref{pro-3}, $r_{i}$ is not a \textbf{Type-4} request.
Hence, by Proposition \ref{costsummary}, the first term in the allocated cost of $r_{i}$ is bounded by $\max\{2, \min\{\gamma, 3\}\}\cdot\lambda$. 
Therefore, it follows 
that 
\begin{eqnarray}
\lefteqn{\textbf{Online$(k+1, i)$}} \nonumber \\ 
& \leq & \max\{2, \min\{\gamma, 3\}\}\cdot\lambda \cdot 
(|Q_1| + |Q_2| + |Q_3| + |Q_5| + |Q_6|) \nonumber \\ & & +\max\{2, \min\{\gamma, 3\}\}\cdot\lambda 
+ \!\!\!\!\!\! \sum_{r_j \in Q_2 \cup Q_3 \cup Q_5 \cup Q_6 \cup \{r_i\}}
\!\!\!\!\!\!\!\!\!\!\!\!\!\!\!\!\!\! \min\{\gamma, 3\} \cdot \mu(s_1)\cdot\big(t_{j}-t'_{j}\big) \nonumber \\
& & + \sum_{r_j \in Q_4}
\mu[r_{j}]\cdot\big(t_{j}-t_{p(j)}\big). 
\label{2C}
\end{eqnarray}

Note that the third and fourth terms above are the storage costs of special copies. We can bound them by using Proposition \ref{pro-2}.

\begin{Proposition} 
It holds that
\begin{equation*}
\sum_{r_j \in Q_2 \cup Q_3 \cup Q_5 \cup Q_6 \cup \{r_i\}}
\big(t_{j}-t'_{j}\big) \leq t_{i}-\Big(t_{h}+\frac{\lambda}{\mu[r_{h}]}\Big).
\end{equation*}
\label{lemma1C}
\end{Proposition}
\begin{proof}
In the online algorithm, according to Proposition \ref{pro-2}, the storage periods $(t'_j, t_j)$ of all the special copies relevant to \textbf{Type-2/3/5/6} requests $r_j \in Q\cup\{r_i\}$ do not overlap. Apparently, all these storage periods end before or at the time $t_i$ of $r_i$. Also note that after serving $t_{h}$, there is a regular copy at server $s[r_{h}]$ over the period $\Big(t_{h}, t_{h}+\frac{\lambda}{\mu[r_{h}]}\Big)$. By Proposition \ref{pro-2} again, the storage period of this regular copy does not overlap with that of any special copy. Thus, the storage periods of all the aforementioned special copies must start no earlier than time $t_{h}+\frac{\lambda}{\mu[r_{h}]}$. 
Hence, the proposition follows.
\end{proof}

It follows from (\ref{2C}) and Proposition \ref{lemma1C} as well as $\min\{\gamma, 3\} \leq \max\{2, \min\{\gamma, 3\}\}$ and $\mu(s_1) \leq \mu[r_h]$ that
\begin{eqnarray*}
\lefteqn{\textbf{Online$(k+1, i)$}} \nonumber \\
& \leq & \max\{2, \min\{\gamma, 3\}\}\cdot\lambda \cdot 
(|Q_1| + |Q_2| + |Q_3| + |Q_5| + |Q_6|) \nonumber \\ & & +\max\{2, \min\{\gamma, 3\}\}\cdot\lambda \nonumber \\
& & + \max\{2, \min\{\gamma, 3\}\}\cdot\mu[r_{h}]\cdot\bigg(t_{i}-\Big(t_{h}+\frac{\lambda}{\mu[r_{h}]}\Big)\bigg) \nonumber \\
& & + \sum_{r_j \in Q_4}
\mu[r_{j}]\cdot\big(t_{j}-t_{p(j)}\big) \nonumber \\
& = & \max\{2, \min\{\gamma, 3\}\}\cdot\lambda \cdot 
(|Q_1| + |Q_2| + |Q_3| + |Q_5| + |Q_6|) \nonumber \\ & & + \max\{2, \min\{\gamma, 3\}\}\cdot\mu[r_{h}]\cdot\big(t_{i}-t_{h}\big) \nonumber \\
& & + \sum_{r_j \in Q_4}
\mu[r_{j}]\cdot\big(t_{j}-t_{p(j)}\big) \nonumber \\
& \leq^{\text{by (\ref{1C})}} & \max\{2, \min\{\gamma, 3\}\} \cdot \textbf{OPT$(k+1,i)$}.
\end{eqnarray*}

\section{\textbf{Type-D} Case}
\label{typed}

$r_{i}$ is a \textbf{Type-D} request in the optimal offline strategy, i.e., $r_{i}$ is served by a transfer from an intermediate server $s_{x}$ whose copy is created by a transfer from a third server when that server receives a request $r_{k}$ (where $k < i$) (Figure \ref{12}(d)). 

\begin{figure}[htbp]
\centering
\includegraphics[width=8cm]{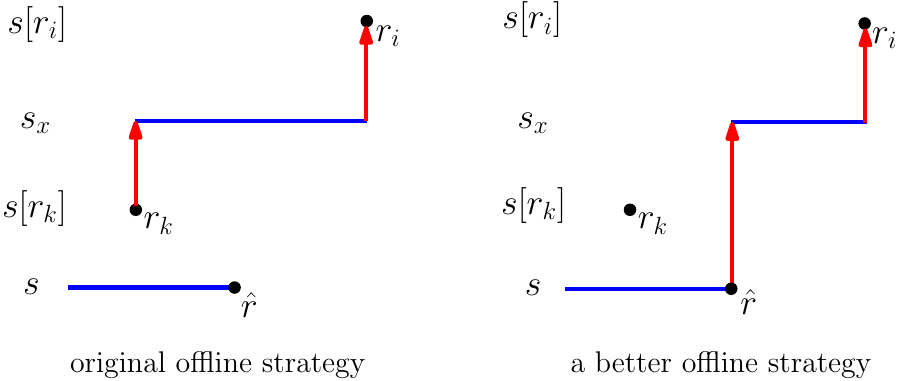}
\caption{Illustration of a data copy crossing time $t_{k}$}
\label{illD}
\end{figure}

In this case, Since $r_i$ is served by a transfer in the optimal offline strategy, by Proposition \ref{prop5}, it must hold that $t_{i}-t_{p(i)}>\frac{\lambda}{\mu[r_i]}$
if $r_{p(i)}$ exists. 
Moreover, it is impossible to have any 
copy that crosses time $t_{k}$. 
This can be proved by contradiction. Assume on the contrary that a copy in some other server $s$ ($s \neq s[r_k]$) crosses $t_k$ in the optimal offline strategy. By similar arguments to Figure \ref{fig_crossing}, the copy must be kept till at least 
the first local request $\hat{r}$ at $s$ after $t_{k}$ 
and will serve $\hat{r}$. By letting $s$ transfer the object to server $s[r_i]$ after serving $\hat{r}$, we can save the storage cost in $s[r_i]$ 
(see Figure \ref{illD}). 
This contradicts that the offline strategy is optimal. 

The rest of the analysis is generally similar to the \textbf{Type-A} case.
We will calculate \textbf{Online$(k+1,i)$} and \textbf{OPT$(k+1,i)$}, and show that $\frac{\textbf{Online$(k+1,i)$}}{\textbf{OPT$(k+1,i)$}} \leq \max\{2, \min\{\gamma, 3\}\}$.
\begin{comment}
Then combining with the induction hypothesis, we have
%can conclude that 
$\frac{\textbf{Online($1,i$)}}{\textbf{OPT($1,i$)}} 
%= \frac{\textbf{Online($1$, $k$)}+\textbf{Online($k+1$, $i$)}}{\textbf{OPT($1$, $k$)}+\textbf{OPT($k+1$, $i$)}} 
\leq \max\{2,\gamma\}$.
\end{comment}

Again, let the set of requests  $Q:=\left\{r_{k+1}, r_{k+2},..., r_{i-1}\right\}$, and divide it into $Q_1 \sim Q_6$ based on the request categorization in the online algorithm, i.e., $Q_i$ includes all the \textbf{Type-$i$} requests in $Q$.

By similar arguments to the \textbf{Type-A} case, the total cost of the data copies and transfers to serve the requests in $Q$ in the optimal offline strategy is
\begin{equation*}
\lambda\cdot 
(|Q_1| + |Q_2| + |Q_3| + |Q_5| + |Q_6|) + \sum_{r_j \in Q_4}
\mu[r_{j}]\cdot\big(t_{j}-t_{p(j)}\big).
\end{equation*}

If we remove these data copies and transfers as well as the copy in $s_x$ during $(t_k, t_i)$ and the transfers from $s[r_k]$ to $s_x$ and from $s_x$ to $s[r_i]$, all the requests $\langle r_{1}, r_{2},..., r_{k}\rangle$ can still be served. Thus,
\begin{equation}
\begin{aligned}
& \textbf{OPT$(k+1,i)$} \geq \lambda\cdot 
(|Q_1| + |Q_2| + |Q_3| + |Q_5| + |Q_6|) +2\lambda \\ & + \mu(s_x)\cdot\big(t_{i}-t_{k}\big) + \sum_{r_j \in Q_4}
\mu[r_{j}]\cdot\left(t_{j}-t_{p(j)}\right).
\label{6D}
\end{aligned}
\end{equation}

Since $t_{i}-t_{p(i)}>\frac{\lambda}{\mu[r_i]}$, based on Proposition \ref{costsummary}, 
\textbf{Online$(k+1,i)$} has the same bound as given in (\ref{2}). 

\begin{Proposition} 
It holds that
\begin{equation*}
\sum_{r_j \in Q_2 \cup Q_3 \cup Q_5 \cup Q_6 \cup \{r_i\}}
\big(t_{j}-t'_{j}\big) \leq t_{i}-t_{k}.
\end{equation*}
\label{lemma2D}
\end{Proposition}
\begin{proof}
The proof is similar to Proposition \ref{lemma1}. 
\end{proof}

It follows from (\ref{2}) and Proposition \ref{lemma2D} as well as $\mu(s_1) \leq \mu(s_x)$ that
\begin{eqnarray*}
\lefteqn{\textbf{Online$(k+1, i)$}} \nonumber \\ 
& \leq & \max\{2, \min\{\gamma, 3\}\}\cdot\lambda \cdot 
(|Q_1| + |Q_2| + |Q_3| + |Q_5| + |Q_6|) \nonumber \\ 
& & + \max\{2, \min\{\gamma, 3\}\}\cdot\lambda 
+ \min\{\gamma, 3\}\cdot\mu(s_x)\cdot\left(t_{i}-t_{k}\right) \nonumber \\
& & + \sum_{r_j \in Q_4}
\mu[r_{j}]\cdot\big(t_{j}-t_{p(j)}\big) \nonumber \\
& \leq^{\text{by (\ref{6D})}} & \max\{2, \min\{\gamma, 3\}\} \cdot \textbf{OPT$(k+1,i)$}. 
\end{eqnarray*}

\end{document}